\documentclass{aa}  

\usepackage[varg]{txfonts}
\usepackage{amssymb}
\usepackage{graphicx}

\usepackage{graphicx}	
\usepackage{mathtools}	
\usepackage{amssymb}	
\usepackage{multicol}        
\usepackage{bm}		
\usepackage{pdflscape}	
\usepackage{xcolor}
\usepackage{placeins}
\usepackage{txfonts}
\begin{document} 

   \title{Pulsar wind nebulae meeting the circumstellar medium of their progenitors}


   \author{D. M.-A.~Meyer, 
          \inst{1}
          Z. Meliani\inst{2}
          \and
          D. F. Torres\inst{1,3,4}\\           
          }

    \institute{Institute of Space Sciences (ICE, CSIC), Campus UAB, Carrer de Can Magrans s/n, 
    08193 Barcelona, Spain\\
    \email{meyer@ice.csic.es}
    \and
    Laboratoire Univers et Théories, Observatoire de Paris, 
    Université PSL, Université de Paris, CNRS, F-92190 Meudon, France
    \and
    Institut d’Estudis Espacials de Catalunya (IEEC), 08034 Barcelona, Spain
    \and
    Institució Catalana de Recerca i Estudis Avançats (ICREA), 08010 Barcelona, Spain    
          }
   \date{}

 
  \abstract
   {
A significative fraction of high-mass stars sail away through the interstellar medium of the galaxies. 
Once they evolved and died via a core-collapse supernova, a magnetized, rotating 
neutron star (a pulsar) is usually their leftover.
The immediate surroundings of the pulsar is
the pulsar  wind, which forms a nebula whose morphology is shaped by the supernova ejecta, channeled  into the circumstellar medium of the progenitor star in the pre-supernova time.     
   }
   {  
Consequently, 
irregular pulsar wind nebulae display a large variety of  
radio appearances, screened by their interacting supernova blast wave and/or 
harboring asymmetric up–down emission.     
   }
   {
Here, we present a series of 2.5-dimensional 
(2 dimensions for the scalar quantities plus a toroidal component for the vectors) 
\textcolor{black}{non-relativistic} magneto-hydrodynamical simulations exploring the 
evolution of the pulsar wind nebulae (PWNe) generated by a red supergiant and a Wolf-Rayet massive 
supernova progenitors, moving with Mach number $M=1$ and $M=2$ into the warm phase of 
the galactic plane. \textcolor{black}{In such a simplified approach, the progenitor's direction of motion, 
the local ambient medium magnetic field, the progenitor and pulsar axis of rotation, are all aligned, 
which restrict our study to peculiar pulsar wind nebula of high equatorial energy flux.}  
   }
   {
We found that the reverberation of the termination shock of the pulsar wind nebulae, 
when sufficiently embedded into its dead stellar surroundings and interacting with 
the supernova ejecta, is asymmetric and differs greatly as a function of the past 
circumstellar evolution of its progenitor, which reflects into their projected 
radio synchrotron emission. 
This mechanism is particularly at work in the context of 
remnants involving slowly-moving and/or very massive stars.    
   }
   {
We find that the mixing of material in plerionic core-collapse supernova 
remnants is strongly affected by the asymmetric reverberation in their pulsar wind 
nebulae.   
   }


   \maketitle
%


\section{Introduction}
\label{intro}

Pulsars are the leftovers of dying massive stars which gravitationally collapse after having
consumed their nuclear fuel \citep{baade_46_phrv_1934}. Conservation of angular momentum
in the last seconds of massive star evolution provides neutron stars with their peculiar
rotational and magnetic properties, potentially spinning up to millisecond frequencies
and exhibiting a 
magnetic field strength reaching up to $10^{15}\, \mathrm{G}$ \citep{manchester_aj_129_2005,lorimer_lrr_2008}. 
The formation of neutron stars is accompanied by the production of gravitational waves, when 
the supernova is asymmetric 
\citep{taylor_apj_253_1982,2013CRPhy..14..318K} and is closely related to the production
of neutrinos \citep{kuroda_apj_755_2012,kotake_2012_AdAst_2012, gabler_mnras_502_2021,2023arXiv230905161S}. 
The pulsar might receive a natal kick,
potentially 
accelerating it to speeds of the order of $100$ to $1000\, \mathrm{km\, s^{-1}}$ 
\citep{Pavan_etal_2016A&A...591A..91P,verbunt_aa_608_2017,vries_apj_908_2021}.
Once born, the pulsar generates a relativistic wind made of leptons, which is blown into such 
\textit{plerionic} supernova ejecta.
The region filled with this material is called 
pulsar wind nebula (PWN), see e.g., \cite{reynolds_apj_278_1984}.

Plerions are therefore supernova remnants that contain a pulsar wind, an archetypical 
e.g., \citep{Hester_2008ARA&A..46..127H, behler_RPPh_2014, bock_aj_1998, 
popov_2019, 2019A&A...627A.100H}. 
PWNe emits non-thermally at all frequencies, complicating the distinction with the supernova remnant (SNR) in the system
\citep{weiler_aa_70_1978, 
caswell_mnras_187_1979, weiler_aa_90_1980}. 
This emission relates to 
the continuous injection 
of
relativistic particles in the magnetized pulsar wind blown 
by the rotating neutron star (pulsar), 
and is affected by their 
dynamics, see, e.g., \cite{Atoyan1996, 2009ApJ...703.2051G,Bucciantini2011,Martin2012,Torres2014,Torres2022}.

Pulsar winds act as an additional component influencing the generic process of supernova 
blast wave propagating into the interstellar medium (ISM) of galaxies~\citep{chevalier_apj_258_1982}, 
itself impacted by the circumstellar medium shaped by the massive progenitor 
stars of type II, core-collapse, supernovae~\citep{chevalier_apj_344_1989, 
orlando_aa_622_2019, orlando_aa_645_2021, orland_aa_636_2020, 2022arXiv220201643O, orlando_aa_666_2022}.
The pulsar wind
first propagates into the freely-expanding explosion material before interacting
with the termination shock of the supernova ejecta, producing a pulsar bubble 
developing strong Rayleigh-Taylor instabilities at the pulsar wind/supernova ejecta contact discontinuity~\citep{reynolds_apj_278_1984, blondin_apj_563_2001, vanderswaluw_aa_404_2003}.
Specifically, the interaction of the supernova blast wave with the circumstellar medium 
is responsible for a variety of observed features, e.g., in the light curves of the growing remnants~\citep{soker_apj_907_2021, soker_raa_2023, soker_raa_2023_23l1001S, shishkin_mnras_522_2023}.
The shocked pulsar wind material 
changes at different stages of the further evolution,  by adopting the morphology of a bow 
shock under the influence of the 
neutron star intrinsic motion~\citep{kulkarni_natur_335_1988}, by developing kinked jet-like features as the 
neutron star spins fast~\citep{weisskopf_apj_765_2013}, or by having its termination 
shock reverberated 
towards the pulsar~\citep{bandiera_mnras_508_2021,Bandiera_mnras_165_2023, 2023_mnras_2839_2023}

\textcolor{black}{
The modelling of pulsar wind nebula has been mostly driven by the study of the 
Crab nebula and a few other iconic historical plerions such as Geminga and Vela, 
driving theoretical advances \citep{kennel_apj_283_1984,coroniti_apj_349_1990,
begelman_apj_397_1992,begelman_apj_493_1998}, axisymmetric simulations 
\citep{Del_Zanna_etal_2006A&A...453..621D, Camus_etal_2009MNRAS.400.1241C, 
komissarov_mnras_414_2011, Olmi_etal_2014MNRAS.438.1518O} and full 3D 
models \citep{2013MNRAS.431L..48P,Porth_etal_2014MNRAS.438..278P, 
Olmi_etla_2016JPlPh..82f6301O}. 
The studies of \citet{komissarov_mnras_344L_2003,komissarov_mnras_349_2004, 
zanna_aa_421_2004,komissarov_mnras_367_2006} demonstrate that polar wind nebulae 
produce a diamond-like morphology, made of an equatorial region pushed by the 
magneto-rotational properties of the pulsar magnetosphere plus a polar 
double jet growing perpendicularly to the disc. 
At early times the pulsar wind leaves the surface of the magnetosphere of the 
spinning neutron star and interacts with its immediate environment, see the 
reviews of \citet{Gaensler_Slane_2006ARA&A..44...17G,
Kargaltsev_etal_2017JPlPh..83e6301K, Olmi_Bucciantini_2023PASA...40....7O} for details. 
This explains the multi-wavelength emission of plerions 
predominantly governed by non-thermal synchrotron and inverse Compton radiation, 
see \citet{Atoyan1996,Reynolds_etal_2017SSRv..207..175R}, but also radio 
\citep{Driessen_etal_2018ApJ...860..133D}, 
X-rays and (up to PeV) $\gamma$-rays 
\citep{Borkowski_etal_2016ApJ...819..160B,2006A&A...457..899A, 
Abdall_etal_2021ApJ...917....6A,Lhaaso_Collaboration_2021Sci...373..425L}.
%
Current and forthcoming ground-based high-energy facilities such as \textit{LHAASO} 
and \textit{CTA} are particularly meant for the study of those emissions, for recent discussions see, e.g., 
\citet{2022MNRAS.517.3550M,acero_aph_2023} and references therein.
}

On a larger scale, more global simulations have been performed in order 
to address the question of pulsar wind nebulae as an essential component 
of the internal functioning of older ($>10\, \mathrm{kyr}$) core-collapse 
supernova remnants. 
There, the pulsar wind interacts with the supernova ejecta and reverberates, 
during and after which the morphology, spectrum, and dynamics of PWN could 
undergo significant changes, see the recent studies by \cite{
Bandiera_mnras_499_2020,Bandiera_mnras_165_2023,2023_mnras_2839_2023}.
The bulk motion of the kicked pulsar throughout the supernova ejecta and the 
distortion of pulsar wind nebulae into bow shocks are modeled in the studies of 
\citet{swaluw_aa_397_2003,swaluw_aa_420_2004,bucciantini_aa_422_2004,
temim_apj_851_2017,kolb_apj_844_2017,temim_apj_932_2022}, in the frames of 
hydrodynamics, magnetohydrodynamics, and relativistic magnetohydrodynamics. 
Applications to specific objects are numerous, 
see for instance the case of the 
the supernova 
remnant G292.0+1.8 \citep{temim_apj_932_2022}, SNR MSH 15-56 
\citep{temim_apj_851_2017}, and SNR G327.1-1.1 \citep{temim_apj_808_2015}. 
The pulsar will eventually leave the still-expanding supernova remnant and 
pass through the forward shock of the supernova blastwave to finally 
interact with the ISM \citep{Bucciantini_aa_375_2001,bucciantini_mnras_478_2018,
Toropina_etal_2019MNRAS.484.1475T}.

The modeling of supernova remnants of massive runaway stars is motivated 
by two facts.
On the one side, up to half of high-mass stars are fast-moving objects 
sailing through the ISM, therefore the inclusion of stellar motion in the 
models is an essential parameter to account for their understanding.
On the other hand, the interaction mechanism at work when a supernova blastwave collides 
with the dense stellar wind bow shocks those stars produce is more prone to 
induce asymmetric structures than when the pre-supernova circumstellar 
medium is spherically-symmetric \citep{velazquez_apj_649_2006,
chiotellis_mnras_435_2013,meyer_mnras_450_2015}.

\textcolor{black}{So far,}
the effects of the stellar motion on the off-centering of the explosion 
into the stellar wind bubble of a $60\, \mathrm{M}_\odot$ \citep{groh_aa564_2014} 
was examined in the work of \citet{meyer_mnras_493_2020}. 
How the coupling 
of the stellar bulk motion with the mass-loss history of a $35\, \mathrm{M}_\odot$ 
Wolf-Rayet supernova progenitor \citep{ekstroem_aa_537_2012} affects the 
non-thermal radio emission of core-collapse supernova remnants is studied 
in \citet{meyer_mnras_502_2021}, demonstrating that bilateral and horseshoe-like 
remnant morphologies are natural consequences of it. The mixing of materials 
(stellar wind, ejecta) in such supernova remnants has been modeled 
in \citet{meyer_mnras_521_2023}. 
In \citet{meyer_mnras_515_2022,meyer_527_mnras_2024} we have 
simulated the additional incidence of a pulsar wind into the above-described 
core-collapse supernova remnants, showing that the pre-supernova distribution 
of circumstellar material governs the evolution of older pulsar wind nebulae, 
both in the context of a moving \citep{meyer_mnras_515_2022} and static 
\citep{meyer_527_mnras_2024} progenitor star. The present work expands onto 
those simulation methodologies.

%
\textcolor{black}{
Two main assumptions characterise this as well as the precedent works of our present study. 
First the choice of an axisymmetric setup, where direction of motion and 
axis of rotation of the progenitor star, as well as the direction of the local 
ISM magnetic field but also the pulsar axis of rotation are aligned, 
because of the intrinsic nature of the used coordinate system.
This 
restricts us to the production of results applying to peculiar pulsar wind 
nebulae, shaped by their higher equatorial energy flux. 
Hence, our models only apply to a specific part of the huge parameter space 
of the pulsar wind nebula problem. 
We decided to investigate it first 
since the use of axisymmetric simulations allows us to reach high spatial 
resolutionsreach high spatial across large regions of space and covering a long duration. 
Secondly, the plerionic supernova remnants are treated within the non-relativistic 
regime. By carrying out classical simulations and using significantly slower pulsar 
winds (recall that the wind can reach Lorentz factors as high 
as $10^6$ \citep[e.g.,][]{kennel_apj_283_1984}), we have access to a more stable 
simulations setup that is more suitable to explore the parameter space in terms of 
stellar evolution history. 
However, reducing the pulsar wind speed can alter several aspects 
of the pulsar wind nebulae evolution, including compression rates, shock velocities  
and the development of associated instabilities. 
However, it is important to note that 
the momentum of the flow is the primary factor influencing the strength of the shock 
\citep{wilkin_459_apj_1996}. By using a lower wind speed, we aim at providing a preliminary 
estimate for the evolution of pulsar wind nebulae. 
}

\textcolor{black}{
We adopt a low magnetization value of \(\sigma = 10^{-3}\), as described 
by \citet{Rees_Gunn_1974MNRAS.167....1R}, \citet{kennel_apj_283_1984}, 
\citet{2017hsn_book_2159S}, \citet{begelman_apj_397_1992}, and \citet{TORRES_et_al_201431}. 
This choice reflects the assumption that a significant portion of the magnetic 
field energy is converted into kinetic energy within the pulsar wind nebulae.
Recent multi-dimensional simulations, however, have demonstrated that higher 
magnetization values, such as \(\sigma = 0.01\) in 2D \citep[e.g.,][]{komissarov_mnras_344L_2003,komissarov_mnras_349_2004,zanna_aa_421_2004,Del_Zanna_etal_2006A&A...453..621D}, 
and even \(\sigma > 1\) in 3D \citep{Porth_etal_2014MNRAS.438..278P, barkov_mnras_484_2019}, 
are more effective in reproducing the features of pulsar wind nebulae termination shocks.
The magnetization of the pulsar wind remains a topic of ongoing debate, as it 
significantly influences the strength of the PWN termination shock and, consequently, 
particle acceleration processes. Furthermore, the magnetization in the equatorial 
wind zone may decrease due to the annihilation of magnetic stripes, leading to lower 
effective magnetization values \citep{Coroniti_2017ApJ...850..184C}. By opting for 
a low magnetization, similar to the value used by \citet{bucciantini_aa_422_2004}, 
the pulsar wind nebulae in our simulations tends to expand more in the equatorial 
plane, resulting in a stronger termination shock.
}

This study investigates the effects of the circumstellar medium of runaway rotating 
massive stars moving in the magnetized warm phase of the galactic plane on the morphology 
and non-thermal emission properties of the pulsar wind nebulae that they generate once they 
have died and exploded as core-collapse supernova remnants.
A representative parameter space of the identity of supernova progenitors is scanned, using 
initial stellar masses of $20\, \rm M_{\odot}$ (the so-called red supergiant progenitor) and 
$35\, \rm M_{\odot}$ (the so-called Wolf-Rayet progenitors), together with space velocities 
of $20\,\rm km\, \rm s^{-1}$ and $40\,\rm km\, \rm s^{-1}$.
It extends the study of \citet{meyer_mnras_515_2022} to several other progenitor stars, both 
in terms of zero-age main-sequence, i.e., assuming different evolutionary paths and bulk motion 
through the ISM. This results in a changing distribution of the circumstellar material at the 
moment of the supernova explosion, which releases a blast wave channeled into it, accordingly 
influencing the morphology of the pulsar wind nebula.

The paper is organised the following manner: Section~\ref{method} presents the 
utilised numerical methods, the results are detailed in Section~\ref{results} 
and discussed in Section~\ref{discussion}. The conclusions of this study are 
drawn in Section~\ref{conclusion}.


\section{Method}
\label{method}

\begin{table*}
	\centering
	\caption{
	Models in our study. The simulations consider the evolution of a rotating massive star 
    at solar metallicity of mass $M_{\star}$ (in $\rm M_{\odot}$), followed from its zero-age 
    main-sequence time that is moving with velocity $v_{\star}$. The ejecta mass 
    $M_{\star}$ (in $\rm ej$) is also indicated. In each one of the models, the ambient medium 
    is that of the warm phase of the galactic plane $n_{\rm ISM}\approx 0.79\, \rm cm^{-3}$. 
    The different evolutionary phases of the stellar evolution are labelled as MS 
    (main-sequence), RSG (red supergiant), WR (Wolf-Rayet), SN (supernova explosion) 
    and PWN (pulsar wind). 
	}
	\begin{tabular}{lcccr}
	\hline
	${\rm {Model}}$           &    $M_{\star}$\, ($\rm M_{\odot}$)   
                              &    $v_{\star}$ ($\rm\, km\, \rm s^{-1}$)
                              &    $M_{\rm ej}\,$ ($\rm M_{\odot}$)
	                           &    Evolution history     \\ 
	\hline   
	PWN-20Mo-v20kms           &  20 &  20   & 7.28
 &  $\rm MS \rightarrow \rm RSG \rightarrow \rm SN \rightarrow \rm PWN $          \\
	PWN-20Mo-v40kms$^{\rm a}$ &  20 &  40   & 7.28
 & $ \rm MS \rightarrow \rm RSG \rightarrow \rm SN \rightarrow \rm PWN$           \\
	PWN-35Mo-v20kms           &  35 &  20   & 10.12
 &  $\rm MS \rightarrow \rm RSG \rightarrow \rm WR \rightarrow \rm SN \rightarrow \rm PWN $\\
	PWN-35Mo-v40kms           &  35 &  40   & 10.12
 &  $\rm MS \rightarrow \rm RSG \rightarrow \rm WR \rightarrow \rm SN \rightarrow \rm PWN$ \\
	\hline    
    \footnotesize 
    (a)~\citet{meyer_mnras_515_2022}
	\end{tabular}
\label{tab:table1}
\end{table*}

\begin{figure}
        \centering
        \includegraphics[width=0.49\textwidth]{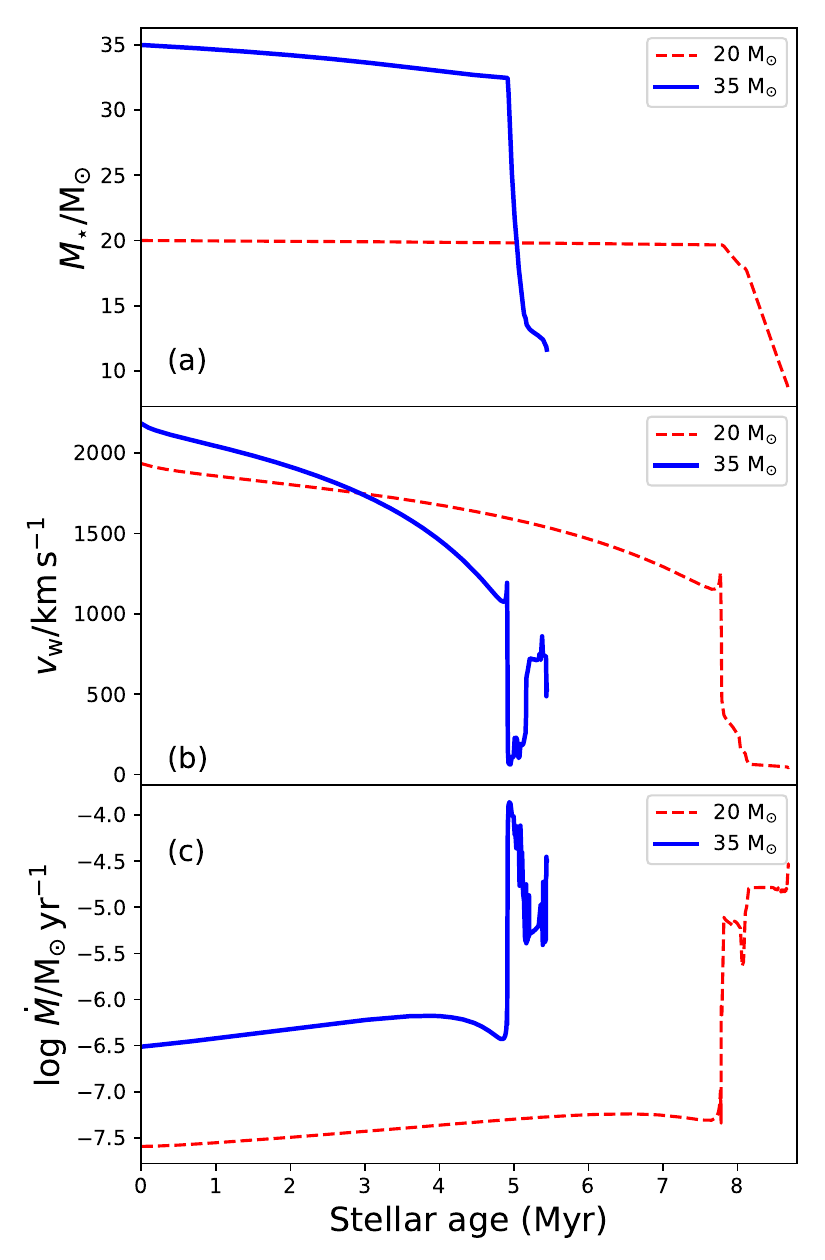}  \\
        \caption{
        Temporal evolution (in $\rm Myr$) of the supernova progenitors of 
        zero-age main-sequence $20\, \rm M_{\odot}$ (dotted red line) 
        and $35\, \rm M_{\odot}$ (solid blue line) considered in our study. 
        The panels display the stellar mass 
        $M_{\star}$ (panel a, in $\rm M_{\odot}$),
        wind velocity $v_{\rm w}$ 
        (panel b, in $\rm km\ \rm s^{-1}$)
        and their mass-loss rate $\dot{M}$ 
        (panel c, in $\rm M_{\odot}\, \rm yr^{-1}$). 
        }
        \label{fig:stellar_properties}  
\end{figure}

In this section
we review the numerical methods, initial conditions and 
physical processes included in the magneto-hydrodynamical simulations, 
as well as the methodology 
used to derive the non-thermal radio emission maps. 

\subsection{\textcolor{black}{Numerical setup}}
\label{strategy}

\subsubsection{\textcolor{black}{Pre-supernova circumstellar medium}}

The wind-ISM interaction of the moving supernova progenitor is calculated 
throughout the entire massive star's life, from its main-sequence 
age
to the pre-supernova 
time. 
We use a \textcolor{black}{cylindrical} computational domain 
$[0;R_{\rm max}]\times[z_{\rm min};z_{\rm max}]$,  where 
$R_{\rm max}=100\, \rm pc$ and $z_{\rm max}=|z_{\rm min}|=50\, \rm pc$,  
which
is mapped with a uniform mesh of $2000 \times 2000$ grid zones. 
\textcolor{black}{
Axisymmetry is imposed along the toroidal direction. 
The $z$ axis is chosen to the direction where the runaway massive 
star moves.
}
Inflow boundary conditions are imposed at $z=z_{\rm max}$ while outflow boundary 
conditions are assigned to $z=z_{\rm min}$ and $R=R_{\rm max}$, respectively, 
since the evolution of the wind-ISM interaction is calculated in the frame of 
the moving star, i.e. by imposing, 
\begin{equation}
    v=-v_\star,
\end{equation}
at $z=z_{\rm max}$ as well as everywhere in the domain, with $v_\star$ the 
velocity of the moving star with respect to the ISM \citep{comeron_aa_338_1998}. 
\textcolor{black}{
The ISM magnetic field is initially everywhere and at $z=z_{\rm max}$ as, 
\begin{equation}
    B = B_\mathrm{ISM},
\end{equation}
with $B_\mathrm{ISM}=7\, \rm \mu G$ is the strength of the magnetic field. The ISM 
magnetic field vector direction is also aligned with the axis of symmetry of the 
computational domain. The ISM density is initially taken to match that of the 
warm phase of the galactic plane $n_{\rm ISM}\approx 0.79\, \rm cm^{-3}$ for a  
temperature of $T_{\rm ISM}\approx 8000\, \rm K$, see \citet{meyer_mnras_515_2022}. 
}

\subsubsection{\textcolor{black}{Supernova remnant and pulsar wind nebula phase}}

\textcolor{black}{
At the supernova time, we remap the wind-ISM solution onto a computational domain of 
smaller size and higher spatial resolution, i.e. a 2.5-dimensional cylindrical 
coordinate system $R_{\rm max}=z_{\rm max}=|z_{\rm min}|=25\, \rm pc$ 
with a uniform mesh of $4000 \times 8000$ grid zones) onto which we inject a 
supernova blastwave followed by a pulsar wind into the computational domain, 
at the location of the defunct runaway star. The spatial resolution of the 
wind-ISM grid is therefore $0.05\, \rm pc$, while that of the pulsar wind 
nebulae grid is $0.00625\, \rm pc$, respectively. 
}
The simulation methodology is that typically used for the modelling of middle-age 
supernova remnants. The entire circumstellar medium of the supernova progenitor 
is first calculated accounting its stellar evolution history. In a second step, 
the stellar wind bow shock is used to model the interaction between the supernova 
blastwave and its surrounding medium. Last, a pulsar wind is injected at the 
location of the ejecta. Hence, one obtains a consistent picture of the evolution 
of the pulsar wind nebula, governed by the distribution of supernova ejecta, 
themselves rules by the circumstellar medium 
\citep{meyer_mnras_515_2022,meyer_527_mnras_2024}.

\subsubsection{\textcolor{black}{Numerical methods}}

The simulations are carried out using the {\sc pluto} 
code~\citep{mignone_apj_170_2007,migmone_apjs_198_2012}\footnote{http://plutocode.ph.unito.it/}. 
\textcolor{black}{ 
The adopted numerical scheme uses a finite-volume formulation with a Godunov-type 
solver made of Harten-Lax-van Leer (HLL) Rieman solver~\citep{hll_ref} for 
flux computation, parabolic reconstruction for spatial accuracy, minmod limiter 
for flux limiting, third-order Runge-Kutta (RK3) for time-stepping, and the 
Courant–Friedrich–Levy condition, initially set to $C_{\rm cfl} = 0.1$ 
for determining the time step. 
}
The use of the eight-wave algorithm~\citep{Powell1997} insures
that $\vec{ \nabla } \cdot  \vec{B} = 0$ all over the calculation domain.

\subsection{Governing equations}
\label{method_eq}


Plasma magneto-hydrodynamics obeys the standard equations 
for the conservation of density,
\begin{equation}
    \frac{\partial \rho}{\partial t} + 
    \boldsymbol{\nabla} \cdot (\rho\mathbf{v}) = 0,
    \label{eq:mhdeq_1}
\end{equation}
linear momentum vector,
\begin{equation}
    \frac{\partial \mathbf{m}}{\partial t} +
    \boldsymbol{\nabla} \cdot \left( \mathbf{m} \otimes \mathbf{v}  
    - \mathbf{B} \otimes \mathbf{B} + \hat{\mathbf{I}}p_{\rm t} \right)
    = \mathbf{0},
    \label{eq:mhdeq_2}
\end{equation}
total energy,
\begin{equation}
    \frac{\partial E}{\partial t} + 
    \boldsymbol{\nabla} \cdot \left( (E+p_{\rm t})\mathbf{v}-\mathbf{B}(\mathbf{v}\cdot\mathbf{B}) \right)
    = \Phi(T,\rho),
    \label{eq:mhdeq_3}
\end{equation}
and magnetic field,
\begin{equation}
    \frac{\partial \mathbf{B}}{\partial t} + 
    \boldsymbol{\nabla} \cdot \left( \mathbf{v} \otimes \mathbf{B} - \mathbf{B} \otimes \mathbf{v} \right)
    = \mathbf{0},
    \label{eq:mhdeq_4}
\end{equation}
where $\rho$ is the mass density, $\mathbf{v}$ the velocity vector,
$\mathbf{m}=\rho\mathbf{v}$ the momentum vector, $\hat{\mathbf{I}}$ 
the identity vector, $\mathbf{B}$ the magnetic field vector, 
$p_{\rm t}=p+\mathbf{B}^{2}/8\pi$ the total pressure\footnote{
\textcolor{black}{
Note that the {\sc pluto} code integrates the equations by absorbing a 
$1/\sqrt{4\pi}$ in the definition of the magnetic field. 
}
}, and, 
\begin{equation}
    E = \frac{p}{(\gamma - 1)} + \frac{\mathbf{m} \cdot \mathbf{m}}{2\rho} + \frac{\mathbf{B} \cdot \mathbf{B}}{2},
    \label{eq:energy}
\end{equation}
the total energy, with $\gamma=5/3$ the adiabatic index. 
\textcolor{black}{
The system is evolved under the assumption of an ideal equation of state and 
the governing equations are closed with,
\begin{equation}
	  c_{\rm s} = \sqrt{ \frac{\gamma p}{\rho} },
\end{equation}
where $c_{\rm s}$ is the sound speed of the plasma. 
}

The right-hand side 
of the 
energy conservation equation is,
%
\begin{equation}  
    \Phi(T,\rho) = n_{\mathrm{H}}\Gamma(T) - n^{2}_{\mathrm{H}}\Lambda(T),
    \label{eq:dissipation}
\end{equation}
where $\Gamma(T)$ and $\Lambda(T)$ are the heating and cooling rates, respectively,
\begin{equation}
    T = \mu \frac{m_{\mathrm{H}}}{k_{\rm{B}}} \frac{p}{\rho},
    \label{eq:temperature}
\end{equation}
the gas temperature, and $n_{\mathrm{H}}$ the hydrogen number density, 
$k_{\rm{B}}$ the Boltzmann constant, $\mu$ the mean molecular weight, 
and $n_{\mathrm{H}}=\rho/(\mu m_{\mathrm{H}} (1+\chi_{\rm He,Z}))$, 
with $\chi_{\rm He,Z}$ the mass fraction of the species heavier than H. 
The heating and cooling laws for ionized gases have been derived in 
\citet{meyer_2014bb} 
\textcolor{black}{
and it is used in the calculation for the pre-supernova remnant phase and the 
early supernova blastwave-stellar wind interaction. This terms are switched-off 
once the pulsar wind is launched and when the system is evolved adiabatically.  
}
%

\subsection{Magnetic stellar wind}
\label{method_wind}

A stellar wind density profile is set onto a sphere of radius equal to 20 grid zones 
centered in the origin, 
\begin{equation}
	\rho_{w}(r,t) = \frac{ \dot{M}(t) }{ 4\pi r^{2} v_{\rm w}(t) }, 
    \label{eq:wind}
\end{equation}
where $\dot{M}(t)$ is the time-dependent mass-loss rate of the progenitor 
star, interpolated from the stellar evolutionary tracks of the {\sc geneva} 
library~\citep{ekstroem_aa_537_2012}. The wind terminal velocity is 
calculated using the recipe of~\citet{eldridge_mnras_367_2006}. 
In Fig. \ref{fig:stellar_properties} we report the time evolution of the 
stellar mass, wind velocity and mass-loss rate used in this study. 
The supernova progenitors are selected with initial rotation at the 
equator such that, 
\begin{equation}
    \frac{ \Omega_{\star}(t=0)}{\Omega_{\rm K}}=0.1,
\end{equation}
with $\Omega_{\star}(t=0)$ the rotational velocities and $\Omega_{\rm K}$ 
the Keplerian velocity at the equator, respectively. The toroidal component 
of the wind velocity therefore reads, 
\begin{equation}
	v_{\phi}(\theta,t) = v_{\rm rot}(t) \sin( \theta ),
\label{eq:Vphi}
\end{equation}
with $\theta$ the azimuthal angle, such that,   
\begin{equation}    
     \Omega_{\star}(t) = \frac{ v_{\rm rot}(t) }{ R_{\star}(t) }, 
\end{equation}
see~\citet{chevalier_apj_421_1994,rozyczka_apj_469_1996}. 
The magnetic field in the stellar wind is assumed to be a Parker spiral, of 
radial component, 
\begin{equation}
	B_{\rm r}(r,t) = B_{\star}(t) \Big( \frac{R_{\star}(t)}{r} \Big)^{2},
    \label{eq:Br}
\end{equation}
and of toroidal component, 
\begin{equation}
	B_{\phi}(r,t) = B_{\rm r}(r,t) 
	\Big( \frac{ v_{\phi}(\theta,t) }{ v_{\rm w}(t) } \Big) 
	\Big( \frac{ r }{ R_{\star}(t) }-1 \Big),
    \label{eq:Bphi}
\end{equation}
with surface magnetic field strength set to $B_{\star}=500$, $0.2$, 
$=100\, \rm G$ for the main-sequence, red supergiant and Wolf-Rayet phases, 
respectively, see~\citet{vlemmings_aa_394_2002,vlemmings_aa_434_2005,
fossati_aa_574_2015,castro_aa_581_2015,przybilla_aa_587_2016,
castro_aa_597_2017,kervella_aa_609_2018}, that is scaled to the 
strength of the solar magnetic field~\citep{scherer_mnras_493_2020,
herbst_apj_897_2020,baalmann_aa_634_2020,baalmann_aa_650_2021,meyer_mnras_506_2021}.

\subsection{Supernova blastwave}
\label{method_snr}

The core-collapse supernova blastwave is imposed using the method developed 
by~\citet{truelove_apjs_120_1999} and~\citet{whalen_apj_682_2008} 
although see the discussion \cite{bandiera_mnras_508_2021} particularly in what refers to the $n$ parameter. 
It consists 
of imposing the following profiles for the mass density in the $r \leq r_{\rm max}$ 
region of the computational domain:
\begin{equation}
\rho(r) = \begin{cases}
        \rho_{\rm core}(r) & \text{if $r \leq r_{\rm core}$},               \\
        \rho_{\rm max}(r)  & \text{if $r_{\rm core} < r < r_{\rm max}$},    \\
        \end{cases}
	\label{cases}
\end{equation}
where, 
\begin{equation}
   \rho_{\rm core}(r) =  \frac{1}{ 4 \pi n } \frac{ (10 E_{\rm ej}^{n-5})^{-3/2}
 }{  (3 M_{\rm ej}^{n-3})^{-5/2}  } \frac{ 1}{t_{\rm max}^{3} },
   \label{sn:density_1}
\end{equation}
and, 
\begin{equation}
   \rho_{\rm max}(r) =  \frac{1}{ 4 \pi n } \frac{ \left(10 E_{\rm
ej}^{n-5}\right)^{(n-3)/2}  }{  \left(3 M_{\rm ej}^{n-3}\right)^{(n-5)/2}  } \frac{ 1}{t_{\rm max}^{3} } 
\bigg(\frac{r}{t_{\rm max}}\bigg)^{-n},
   \label{sn:density_2}
\end{equation}
with $M_{\rm ej}$ representing the ejecta mass and $E_{\rm ej}$ the ejecta energy, 
$n=11$ being an exponent typical for core-collapse supernova 
blastwaves~\citep{truelove_apjs_120_1999}\footnote{We note that this brings additional uncertainty, see the discussion 
in \cite{bandiera_mnras_508_2021}, especially associated with their figure 4. We hope to incorporate this description in the future.} 
Also, 
\begin{equation}
    t_{\rm max} = \frac{r_{\rm max}}{v_{\rm max}},
\end{equation}
which is determined by the numerical iterative method described 
in~\citet{whalen_apj_682_2008}.

The velocity field is set radially, using the relation, 
\begin{equation}
    v(r) = \frac{ r }{ t },
\end{equation}
with, 
\begin{equation}
   v_{\mathrm{core}} = \bigg(  \frac{ 10(n-5)E_{\mathrm{ej}} }{ 3(n-3)M_{\mathrm{ej}} } \bigg)^{1/2},
   \label{sn:vcore}
\end{equation}
and, 
\begin{equation}
v_{\mathrm{max}}= \frac{r_{\mathrm{max}}}{t_{\mathrm{max}}}=3\times 10^{4}\, \mathrm{km\, s^{-1}},
\end{equation}
at the characteristic radii $r_{\mathrm{core}}$ and $r_{\mathrm{max}}$, respectively, 
marking the transition between the plateau and the forward shock of the blastwave. 
The mass in the ejecta is chosen to be that of the progenitor at the moment 
of the explosion, minus the mass of the neutron star that is left behind, 
\begin{equation}
   M_{\mathrm{ej}} =  M_{\star} - \int_{t_\mathrm{ZAMS}}^{t_\mathrm{SN}} \dot{M}(t)~ dt - M_{\mathrm{NS}},
   \label{eq:co}
\end{equation}
with $M_{\mathrm{NS}}=1.4\, \mathrm{M}_\odot$, and with $t_\mathrm{ZAMS}$ and $t_\mathrm{SN}$ 
the zero-age main-sequence and supernova times, respectively. The ejecta masses in the 
different models are reported in Table \ref{tab:table1}.

\subsection{Pulsar wind}
\label{method_psr}

The pulsar wind 
description
is based on the method developed in~\citet{komissarov_mnras_349_2004}. 
It is controlled by its time-dependent total power,
\begin{equation}
\dot{E}(t) = \dot{E}_{\mathrm{o}} \left(1 + \frac{t}{\tau_{\mathrm{o}}} \right)^{\alpha},
\end{equation}
\textcolor{black}{
with,  
\begin{equation}
\alpha = \frac{  n+1 }{  n-1 },
\end{equation}
where n is the braking index of the pulsar,} 
and with $\dot{E}_{\mathrm{o}} = 10^{38}\,\mathrm{erg}\,\mathrm{s}^{-1}$ 
as the initial pulsar energy variation, and the initial spin-down given by, 
\begin{equation}
\tau_{\mathrm{o}} = \frac{P_{\mathrm{o}}}{(n-1)\dot{P}_{\mathrm{o}}},
\end{equation}
\textcolor{black}{
where following \citet{2017hsn_book_2159S}, we set $n=3$ for the 
pulsar's magnetic dipole spin-down.}  
The pulsar's initial period is set 
to $P_{\mathrm{o}} = 0.3\,\mathrm{s}$ and its time-derivative is 
$\dot{P}_{\mathrm{o}} = 10^{-17}\,\mathrm{s}\,\mathrm{s}^{-1}$.
The model assumes a constant pulsar wind velocity,
\begin{equation}
v_{\mathrm{psr}} = 10^{-2} c,
\end{equation}
with $c$ being the speed of light. The mass-loss rate in the pulsar wind depends 
on its total power and velocity, given by, 
\begin{equation}
\dot{M}_{\mathrm{psr}}(t) = \frac{2\dot{E}(t)}{v_{\mathrm{psr}}^{2}},
\end{equation}
therefore its mass density is, 
\begin{equation}
\rho_{\mathrm{psr}}(r,t) = \frac{\dot{M}_{\mathrm{psr}}(t)}{4\pi r^{2}v_{\mathrm{psr}}},
\end{equation}
respectively. The pulsar wind magnetization is exclusively toroidal, with a 
magnetisation factor $\sigma=10^{-3}$ \citep{2017hsn_book_2159S}, so that, 
\textcolor{black}{
\begin{equation}
B_{\mathrm{psr}}(r,t) = 
\sqrt{ 4\pi \frac{ \dot{E}(t) }{ v_{ \mathrm{psr} }  }}  
\frac{\sqrt{  \sigma }}{r}  
\sin(\theta)\left(1 - \frac{2\theta}{\pi}\right),
\end{equation}
}
\textcolor{black}{is the pulsar wind magnetisation at the inner pulsar wind radius $r_{\rm in}$ of 
20 grid zones (of the finer grid $4000 \times 8000$), within which the pulsar wind 
is continuously enforced. There is no evolution of the equations inside of the 
inner boundary of radius $r_{\rm in}$, that is 8 times smaller duing the pulsar 
wind phase, than during the wind-ISM phase. This remapping of the wind-ISM solution 
onto a finer grid at the pre-supernova time, prior to the launching of the pulsar 
wind, is necessary to avoid problems due to reverberation with the termination shock, 
potentially hitting the inner boundary. 
}
A passive scalar tracer $Q_{\mathrm{psr}}=1$ is injected into the pulsar wind with 
an initial value of $Q_{\mathrm{psr}}=0$ elsewhere. It obeys the advection equation,
\begin{equation}
\frac{\partial (\rho Q_{\mathrm{psr}})}{\partial t} + \nabla \cdot (\mathbf{v} \rho Q_{\mathrm{psr}}) = 0,
\end{equation}
which allows the time-dependent tracking of the transport of leptonic material into 
the supernova ejecta and the defunct circumstellar medium.

\subsection{Radiative transfer calculations}
\label{method_rad}

We generate emission maps from the magneto-hydrodynamical simulations. The three-dimensional 
structure of the gas density and magnetic field in the pulsar wind nebula are reconstructed 
from the 2.5-dimensional models accounting for their axisymmetric 
nature~\citep{meyer_mnras_502_2021}. 
We present the maps in normalised units, i.e. relative arbitrary scale, which is done by  
assuming the non-thermal electron population to be described by the power-law distribution, 
\begin{equation}
    N(E) = K E^{-s},
    \label{eq:N}  
\end{equation}
where $K$ is a proportionality constant and $s=2\alpha+1$ is an index that is a 
function of the spectral index $\alpha$. 
The synchrotron emissivity is given by, 
\begin{equation}
    j_{\mathrm{sync}}(\nu,\theta_{\mathrm{obs}}) \propto n B_{\perp}^{(s+1)/2} \nu^{-(s-1)/2},
    \label{eq:coeff}  
\end{equation}
with $B_{\perp}$ as the component of the magnetic field that is normal to the line of 
sight and $\nu$ is the emission frequency. The emission maps are obtained by integrating 
the emissivity along the line of sight, 
\begin{equation}
    I = \int j_{\mathrm{sync}}(\theta_{\mathrm{obs}}) dl,
    \label{eq:intensity}  
\end{equation}
according to a given aspect angle $\theta_{\mathrm{obs}}$ where $dl$ is the line-of-sight 
differential length element~\citep{meyer_mnras_515_2022}.


\begin{figure*}
        \centering
        \includegraphics[width=0.7\textwidth]{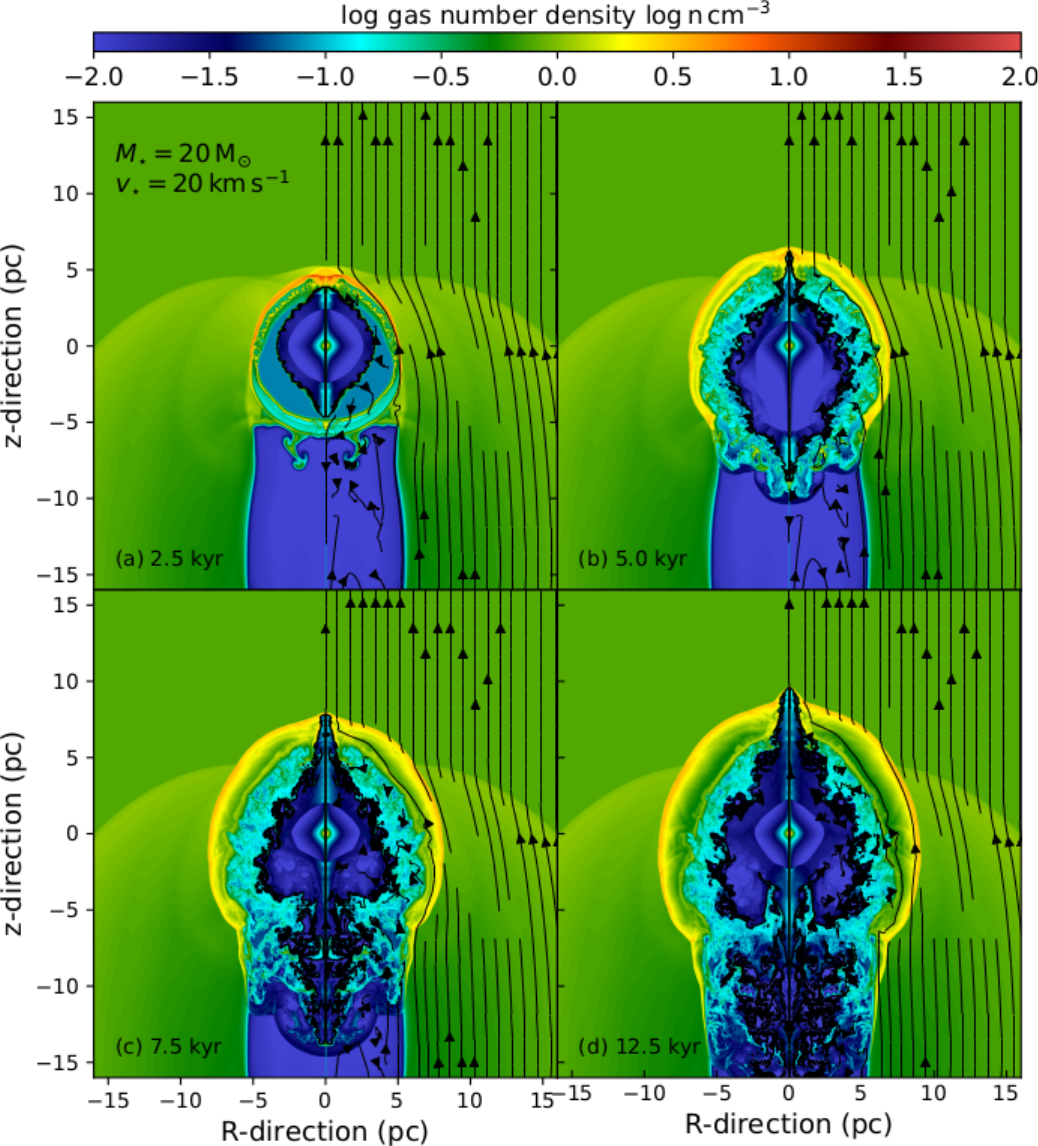}  \\        
        \caption{
         Number density fields in our magneto-hydrodynamical simulation of 
         the supernova remnant of the runaway $20\, \rm M_{\odot}$ star rotating 
         with $\Omega_{\star}/\Omega_{\rm K}=0.1$ and moving with velocity 
         $v_{\star}=20\, \rm km\, \rm s^{-1}$. 
         The evolution of the plerionic supernova remnant is shown at times 
         $2.5$ (a), $5.0$ (b), $7.5$ (c) and $12.5\, \rm kyr$ (d), respectively. 
         The various black contour highlights the region with a $50\%$ per cent  
         contribution of pulsar wind material. 
         The black arrows in the right-hand parts of the figures are  
         ISM magnetic field lines. 
        }
        \label{fig:snr_pwn_2020}  
\end{figure*}

\begin{figure*}
        \centering
        \includegraphics[width=0.7\textwidth]{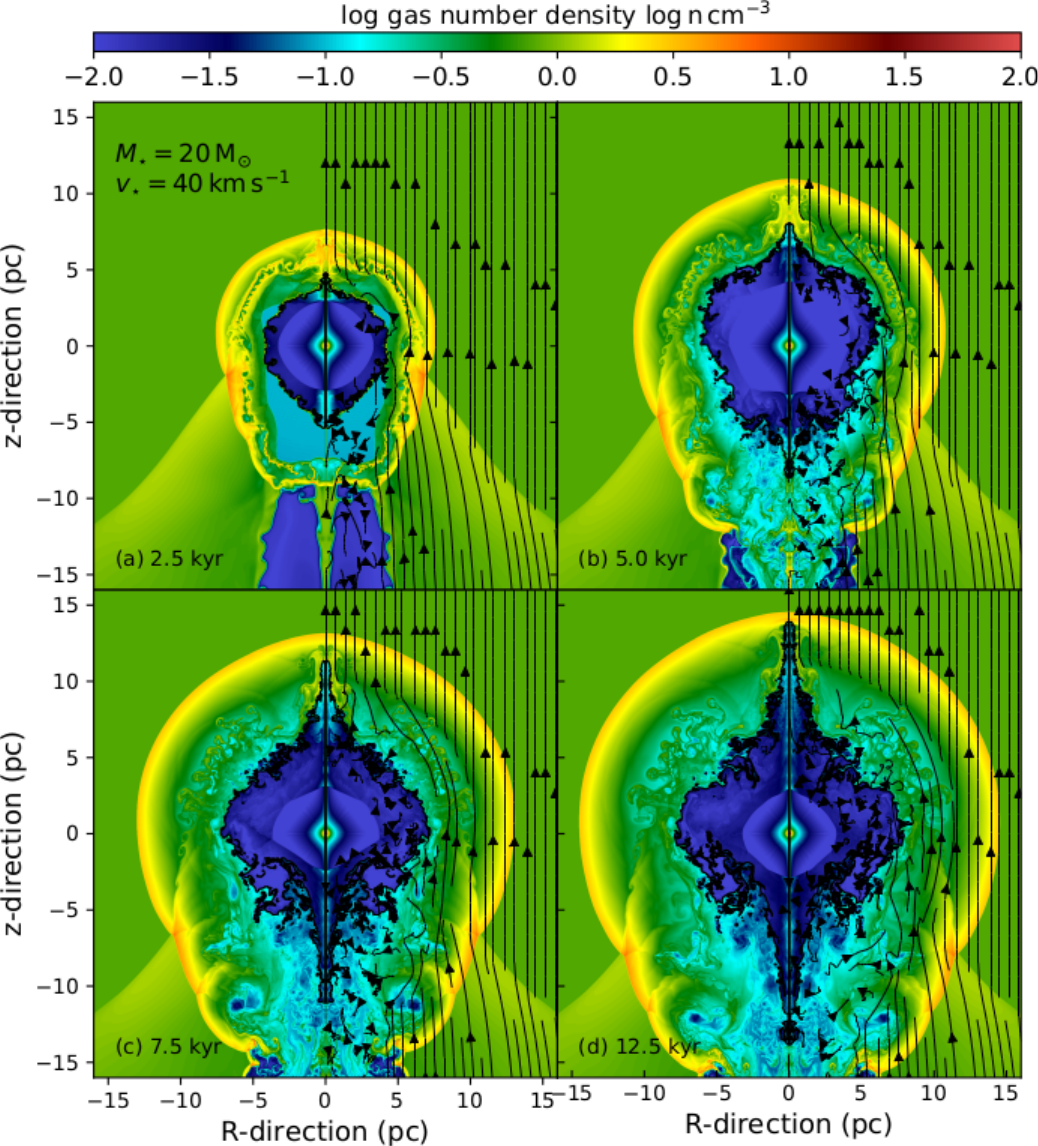}  \\        
        \caption{
        As Fig.~\ref{fig:snr_pwn_2020} for a $20\, \rm M_{\odot}$ progenitor 
        star moving with velocity $v_{\star}=40\, \rm km\, \rm s^{-1}$. 
        }
        \label{fig:snr_pwn_2040}  
\end{figure*}

\section{Results}
\label{results}

\begin{figure*}
        \centering
        \includegraphics[width=0.8\textwidth]{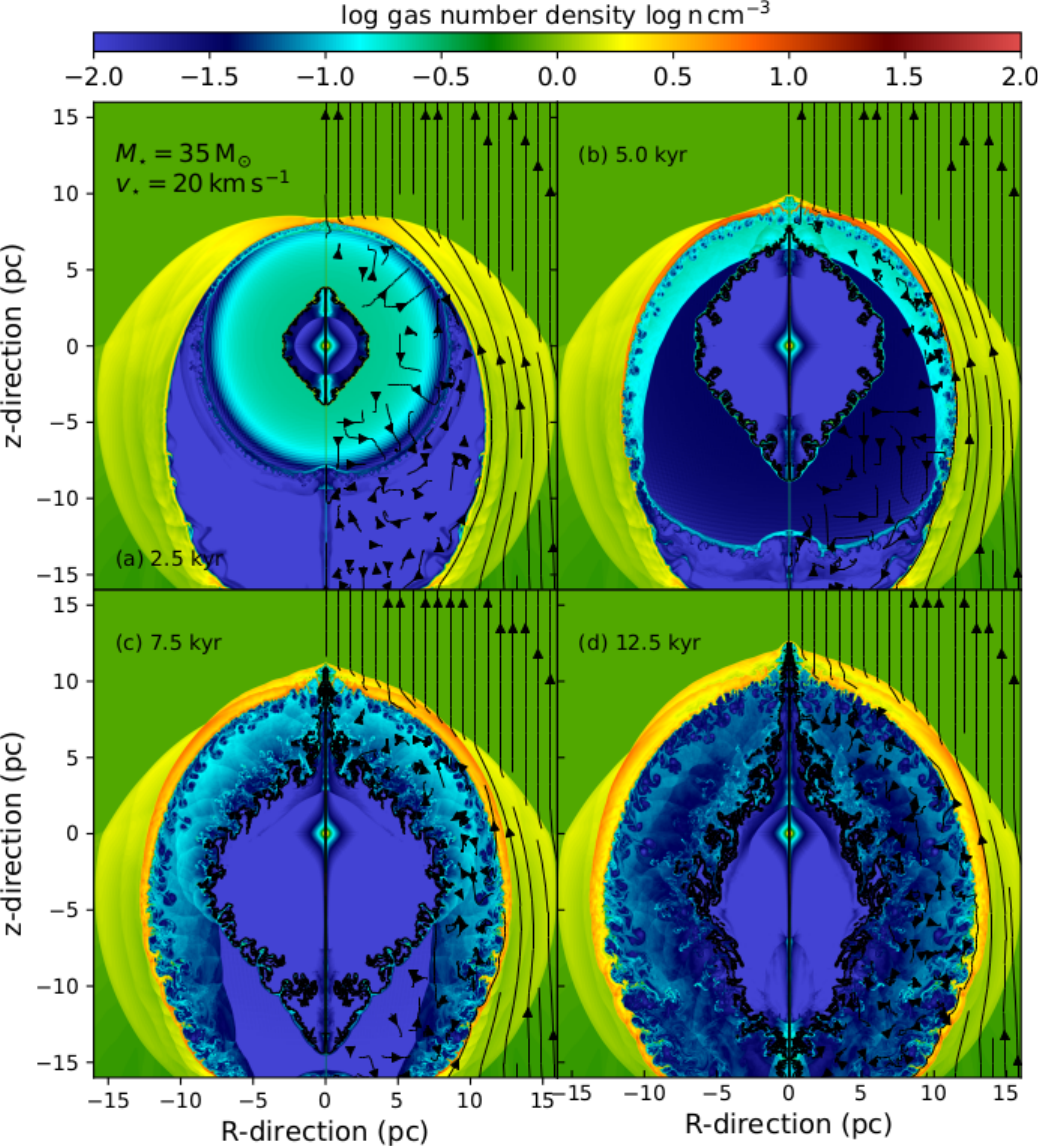}  \\        
        \caption{
        As Fig.~\ref{fig:snr_pwn_2020} for a $35\, \rm M_{\odot}$ progenitor star 
        moving with velocity $v_{\star}=20\, \rm km\, \rm s^{-1}$. 
        }
        \label{fig:snr_pwn_3520}  
\end{figure*}

This section presents the structure of the plerionic supernova remnants in our models 
and describes the non-thermal appearance of the pulsar wind nebulae by synchrotron emission.

\subsection{Model with $20\, \rm M_{\odot}$ and $v_{\star}=20\, \rm km\, \rm s^{-1}$}
\label{resultats_model_1}

Fig.~\ref{fig:snr_pwn_2020} shows the number density fields in the model 
with a runaway red supergiant star of initial mass $20\, \mathrm{M}_{\odot}$, moving 
with velocity $v_{\star}=20\, \mathrm{km\, s^{-1}}$. The panel plots the supernova remnant 
structure at times $2.5$ (a), $5.0$ (b), $7.5$ (c), and $12.5\, \mathrm{kyrs}$ (d) after 
the onset of the supernova explosion, respectively. In the figure, the black 
contour marks the region of the remnant with a $50\%$ contribution of pulsar 
wind material in number density, while the black arrows in the right-hand parts 
of the figures are ISM magnetic field lines.

At 
$2.5\, \mathrm{kyrs}$ the large-scale circumstellar medium occupies most of the 
computational domain, composed of the main-sequence stellar wind bow shock in 
which the red supergiant wind has been blown. 
%
The main-sequence stellar wind bow shock is a structure
within which the region of unshocked wind 
material is shaped as a low-density region that extends behind the dense arc of swept-up 
ISM gas accumulated during the life of the moving star.
The red supergiant wind 
is a shell of slow, 
dense material that collided with the former bow shock and penetrated into it within 
a radius $\approx 10\, \mathrm{pc}$ around the center of the explosion.

Furthermore, the supernova blastwave, launched at the moment of the explosion, has, 
in its turn, interacted with the circumstellar distribution of red supergiant wind. 
This induces large Rayleigh-Taylor instabilities at the wind/ejecta contact 
discontinuity in which the pulsar wind nebula develops, sweeping up the supernova 
ejecta (Fig.~\ref{fig:snr_pwn_2040}a).

At 
$5.0\, \text{kyrs}$, the supernova blastwave is now strongly interacting 
with the red supergiant wind, creating a dense arc of swept-up material made of 
both stellar wind and ambient medium. 
The blastwave travels back towards the center 
of the explosion, creating an additional ovoidal layer filled by a hot mixture of 
evolved stellar wind, ejecta, and pulsar wind materials, which extends up to about 
$4$--$7\, \text{pc}$ from the origin, slightly losing sphericity as it pours out 
into the low-density tail of the bow shock (at $z=-7\, \text{pc}$). 
The shape of 
the pulsar wind nebula deviates from the diamond-like solution of \citet{komissarov_mnras_349_2004}, 
as a direct result of its interaction with the ejecta, 
itself impacted by the 
circumstellar medium of the runaway progenitor \citep{meyer_mnras_515_2022}, see 
Fig. \ref{fig:snr_pwn_2040}b.

At 
$7.5\, \text{kyrs}$ and later, the plerionic supernova remnant has further evolved, 
in the sense that the spinning pulsar induces a double polar jet-like feature, 
either interacting with the bow shock ($z>0$ region of the figure), or expanding 
more freely into the low-density trail of material ($z<0$ region of the figure). 
While the forward shock of the blastwave expands outwards, the power of the rotating 
neutron star sustains the reverberating termination shock of the pulsar wind nebula, 
preventing it from recovering its diamond-like morphology. Consequently, the pulsar 
wind nebula adopts a spinning top shape which is conic in the upper part and jet in 
the lower region (Fig. \ref{fig:snr_pwn_2040}c,d).

\begin{figure*}
        \centering
        \includegraphics[width=0.7\textwidth]{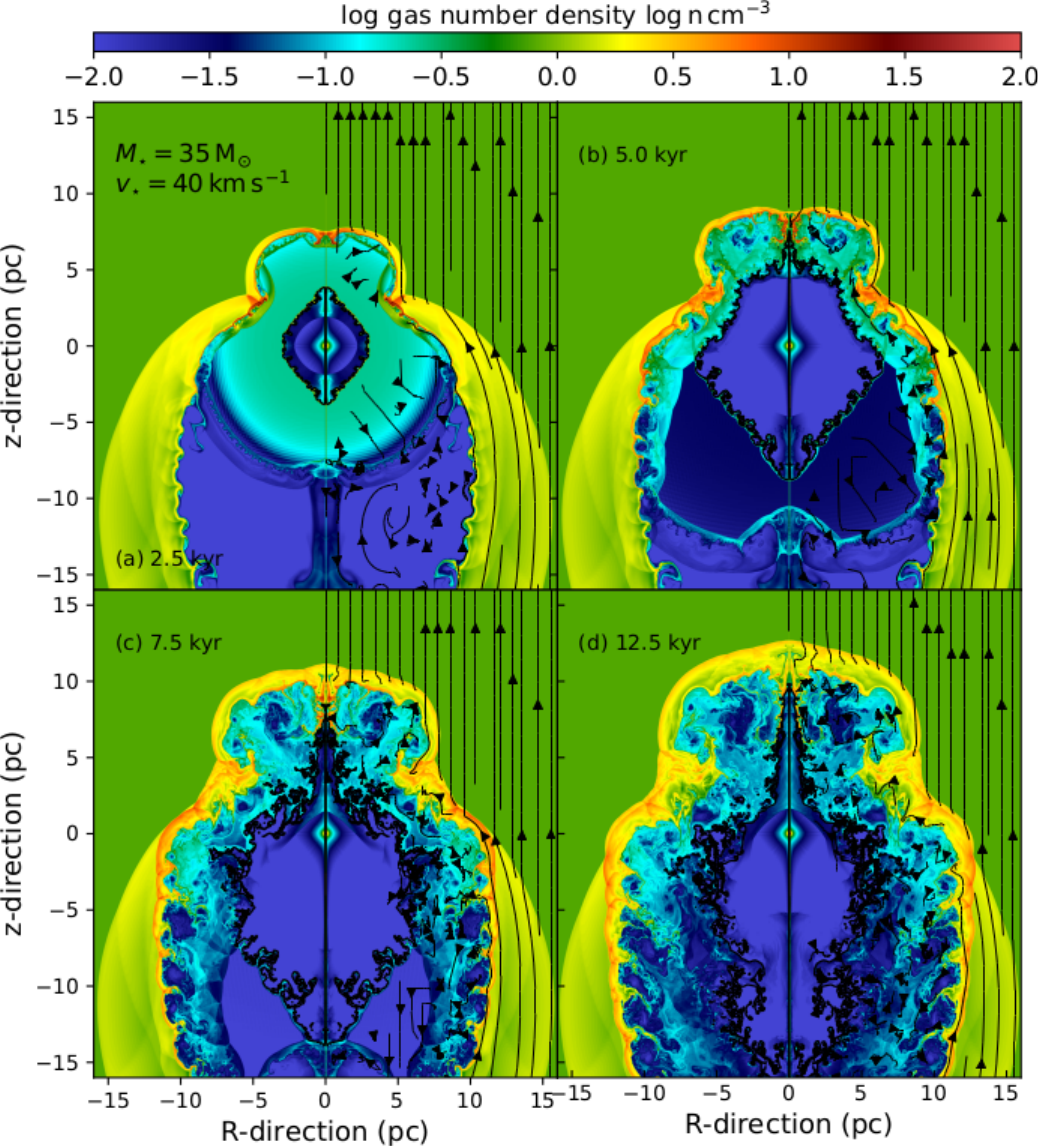}    
        \caption{
        As Fig.~\ref{fig:snr_pwn_2020} for a $35\, \rm M_{\odot}$ progenitor star 
        moving with velocity $v_{\star}=40\, \rm km\, \rm s^{-1}$. 
        }
        \label{fig:snr_pwn_3540}  
\end{figure*}

\subsection{Model with $20\, \rm M_{\odot}$ and $v_{\star}=40\, \rm km\, \rm s^{-1}$}
\label{resultats_model_2}

Fig.~\ref{fig:snr_pwn_2040} shows the number density fields in the simulation 
model with a runaway red supergiant star of initial mass $20\, \mathrm{M_{\odot}}$, 
moving with velocity $v_{\star}=40\, \mathrm{km\, s^{-1}}$, already described 
in great detail in the preceding paper of this series, see \citet{meyer_mnras_515_2022}.
The circumstellar medium generated by the runaway massive star is different 
from that in the model with velocity $v_{\star}=20\, \mathrm{km\, s^{-1}}$. 
Even if the stellar evolutionary stages it undergoes are similar, the morphology 
of the stellar wind bow shock is 
different as a result of the bulk 
motion, producing a smaller stand-off distance \citep{baranov_sovastronlett_1_1975} 
and a more compact bow shock \citep{wilkin_459_apj_1996}.
The main-sequence bow shock is closer than in the model with $v_{\star}=20\, \mathrm{km\, s^{-1}}$ 
and the cavity of low-density material of sensibly more reduced section. The 
red supergiant material is contained in a more reduced region, since the denser 
walls of the bow shock prevent it from penetrating deeply into the region 
of main-sequence material (Fig.~\ref{fig:snr_pwn_2040}a).

The moving progenitor in the model with an initial mass of $20\, \rm M_{\odot}$
and a velocity of $v_{\star}=40\, \rm km\, \rm s^{-1}$ is the typical scenario 
leading to the production of a Cygnus Loop supernova remnant \citep{meyer_mnras_450_2015}, 
with an asymmetric propagation of the shockwave that is channeled by the azimuthal-dependent 
distribution of circumstellar material. 
Along the direction of motion of the progenitor, 
it interacts with the bow shock and further expands into the ISM, while in the opposite 
direction, the blast wave propagates through the bow shock tail (Fig. \ref{fig:snr_pwn_2040}a,b).
The pulsar wind nebula that grows inside this simil to the 
Cygnus Loop prevents the blast wave 
reverberation, while the morphology of the pulsar wind nebula sees its shape 
distances from the diamond-like form of \citet{komissarov_mnras_349_2004}. 
In \citet{meyer_mnras_515_2022}, it is shown that the spatial distribution of 
the region of shocked pulsar wind material, located between the lepton/ejecta 
contact discontinuity, is governed by a competition of the magneto-rotational 
intrinsic properties of the pulsar on the one hand, and the pre-supernova 
circumstellar medium on the other hand. 
First, it becomes 
an ovoid region 
(Fig. \ref{fig:snr_pwn_2040}a), then exhibits important north-south 
asymmetries
(Fig. \ref{fig:snr_pwn_2040}b), to end up with a rather 
standard bipolar jet-plus-equatorial structure (Fig. \ref{fig:snr_pwn_2040}c,d).


\subsection{Model with $35\, \rm M_{\odot}$ and $v_{\star}=20\, \rm km\, \rm s^{-1}$}
\label{resultats_model_3}

Fig. \ref{fig:snr_pwn_3520} shows the number density fields in the simulation 
with a runaway Wolf-Rayet star of initial mass $35\, \rm M_{\odot}$, moving with 
a velocity of $v_{\star}=20\, \rm km\, \rm s^{-1}$ through the ISM. 
The pre-supernova circumstellar medium is made of a large ring of dense Wolf-Rayet 
material distributed at a radius of approximately $7-15\, \rm pc$. The power of this 
last stellar wind has pushed away the material from the previous evolutionary 
phases of the massive star, and it mostly interacts with the bow shock rather than 
with the ISM \citep{brighenti_mnras_273_1995}. 
The supernova blast wave is expanding into this ring as well as the pulsar 
wind material into the ejecta, which has the diamond-like morphology described by 
\citet{komissarov_mnras_349_2004}. Refer to Fig. \ref{fig:snr_pwn_3520}a. 
At 
$5.0\, \rm kyrs$, the supernova ejecta have collided with the Wolf-Rayet 
ring and begin to be travel back towards the center of the explosion. 
A region of shocked ejecta material forms while the blast wave still expands 
downwards towards the tail of the bow shock. 
The contact discontinuity of the pulsar wind nebula interacts with the 
reverberating supernova blast wave and begins to lose the structure described  
in the study by \citet{komissarov_mnras_349_2004}.

At 
$7.5\, \mathrm{kyrs}$, the pulsar wind is strongly interacting with 
the supernova ejecta, moving back to the region of the explosion in the $z>0$ 
region of the supernova remnant. A series of unstable Rayleigh-Taylor layers 
form as a mechanism of double reverberation begins: the termination shock 
of the pulsar wind, in its turn, moves back towards the center of the 
explosion, while the termination shock of the supernova blast wave experiences 
a similar mechanism.
At 
$12.5\, \mathrm{kyrs}$, the two termination shocks have met, giving 
birth to an overall region of mixed shocked ejecta and shocked pulsar wind 
engulfing the unshocked, still streaming pulsar wind material and surrounded 
by the bow shock of the defunct stellar wind. As a consequence of the 
asymmetric distribution of circumstellar material, the reverberations 
are more important in the $z>0$ region of the supernova remnant, producing 
an oblong pulsar wind nebula. On the top of the shock, a jet-like feature 
develops and propagates until the outer border of the supernova remnant and to the ISM.
The low-density region, regardless of the material (wind, ejecta, pulsar wind) 
it is composed of, is much larger in the context of a moving Wolf-Rayet 
supernova progenitor than that of a red supergiant star, and the properties 
of the mixed material change accordingly, as well as the size of 
the pulsar wind nebula.

\subsection{Model with $35\, \rm M_{\odot}$ and $v_{\star}=40\, \rm km\, \rm s^{-1}$}
\label{resultats_model_4}

Fig.~\ref{fig:snr_pwn_3540} shows the number density fields in the simulation 
model with a runaway Wolf-Rayet star of initial mass $35\, \mathrm{M}_{\odot}$, 
moving with velocity $v_{\star}=40\, \mathrm{km\, s}^{-1}$ through the ISM. 
As in the models with initial mass $20\, \mathrm{M}_{\odot}$, the stellar 
wind bow shock of the progenitor star is less open when it moves faster. 
Since the last stellar wind of the Wolf-Rayet phase is blown at a terminal 
speed exceeding that of the progenitor bulk motion, the ring of material 
penetrates directly into the unperturbed ISM, see description of this 
mechanism in \citet{brighenti_mnras_277_1995}. It is the site of huge 
Rayleigh–Taylor instabilities which deform the circumstellar ring. 
The expanding supernova blastwave hits the bow shock and fills the large 
eddies it developed and the typical diamond-like solution for the pulsar 
wind nebula forms into it, see Fig.~\ref{fig:snr_pwn_3540}a. 
At this time, the pulsar wind nebula is protected from information of 
the past evolution of the progenitor star. At time $5.0\, \mathrm{kyr}$, the 
pulsar wind has interacted with the reverberating supernova ejecta 
in the direction of stellar motion, while both the blastwave and pulsar 
wind still propagate freely in the other direction.

At time $7.5\, \rm kyr$ the reverberation of the supernova blastwave 
collides with the pulsar wind, which, in its turn, experiences the 
reverberation process, leaving around a swept-up region of shocked 
ejecta and pulsar wind material that is surrounded by the ring of 
Wolf-Rayet stellar wind. The morphology of the pulsar wind is strongly 
deviating from the diamond-like solution of \citet{komissarov_mnras_349_2004}. 
This trend keeps going at later times, and, $12.5\, \rm kyrs$ after the 
supernova explosion, the pulsar wind has adopted a shape 
similar to that in the model with a star of initial mass $35\, \rm M_{\odot}$  
moving with velocity $v_{\star}=40\, \rm km\, \rm s^{-1}$ through the ISM, 
composed of an oblong region in which the pulsar is off-centered and a 
bipolar jet that intercept the dense former stellar wind bow shock hit 
by the supernova blastwave (Fig. \ref{fig:snr_pwn_3540}d).

\subsection{Radio synchrotron maps}
\label{resultats_maps}

\begin{figure*}
        \centering
        \includegraphics[width=0.9\textwidth]{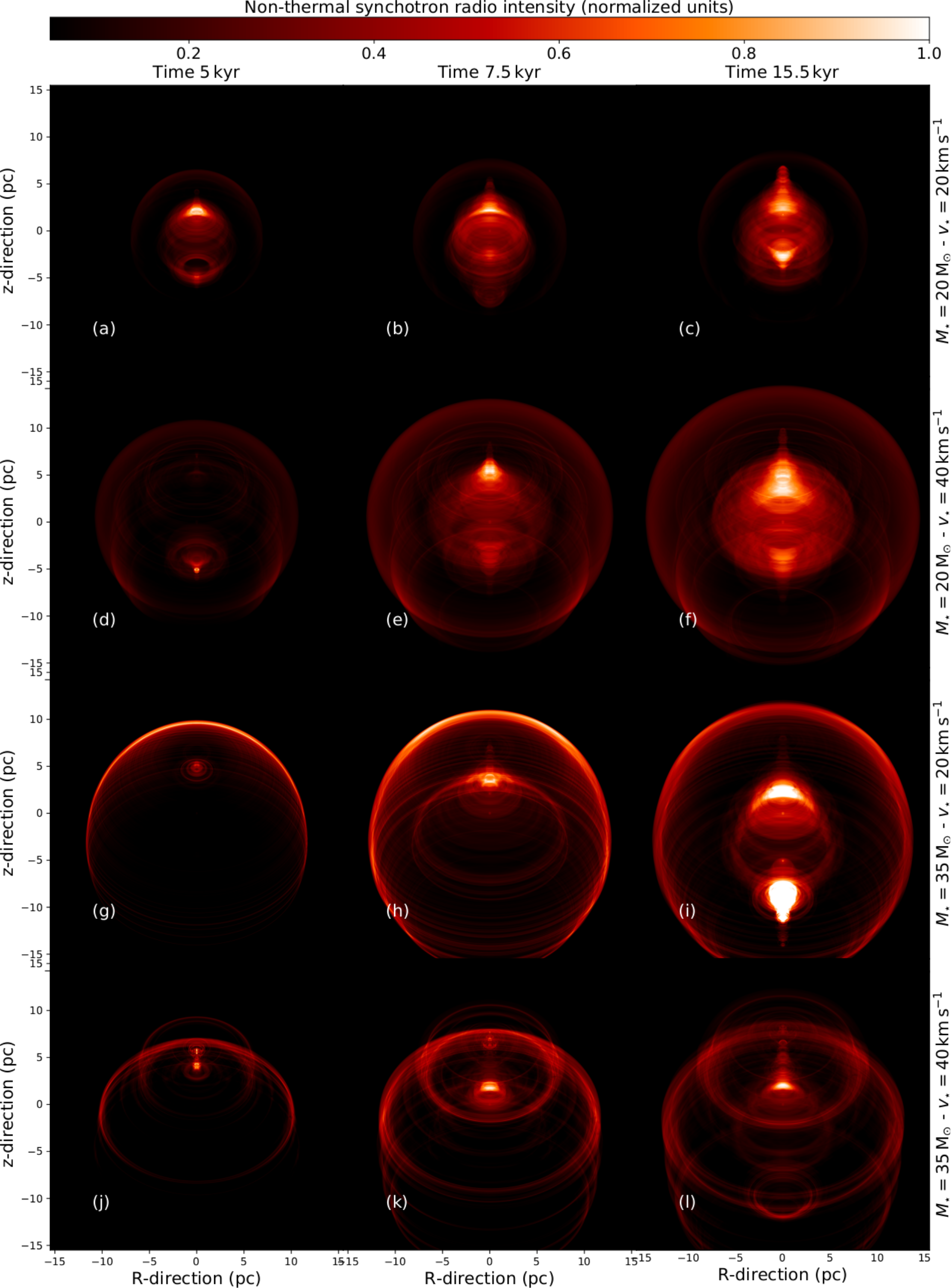}  \\ 
        \caption{
        Synchrotron radio emission maps of the grid of models with 
        $\theta_\mathrm{obs}=45^{\circ}$. 
        }
        \label{fig:maps_sync_45deg}  
\end{figure*}

\begin{figure*}
        \centering
        \includegraphics[width=0.9\textwidth]{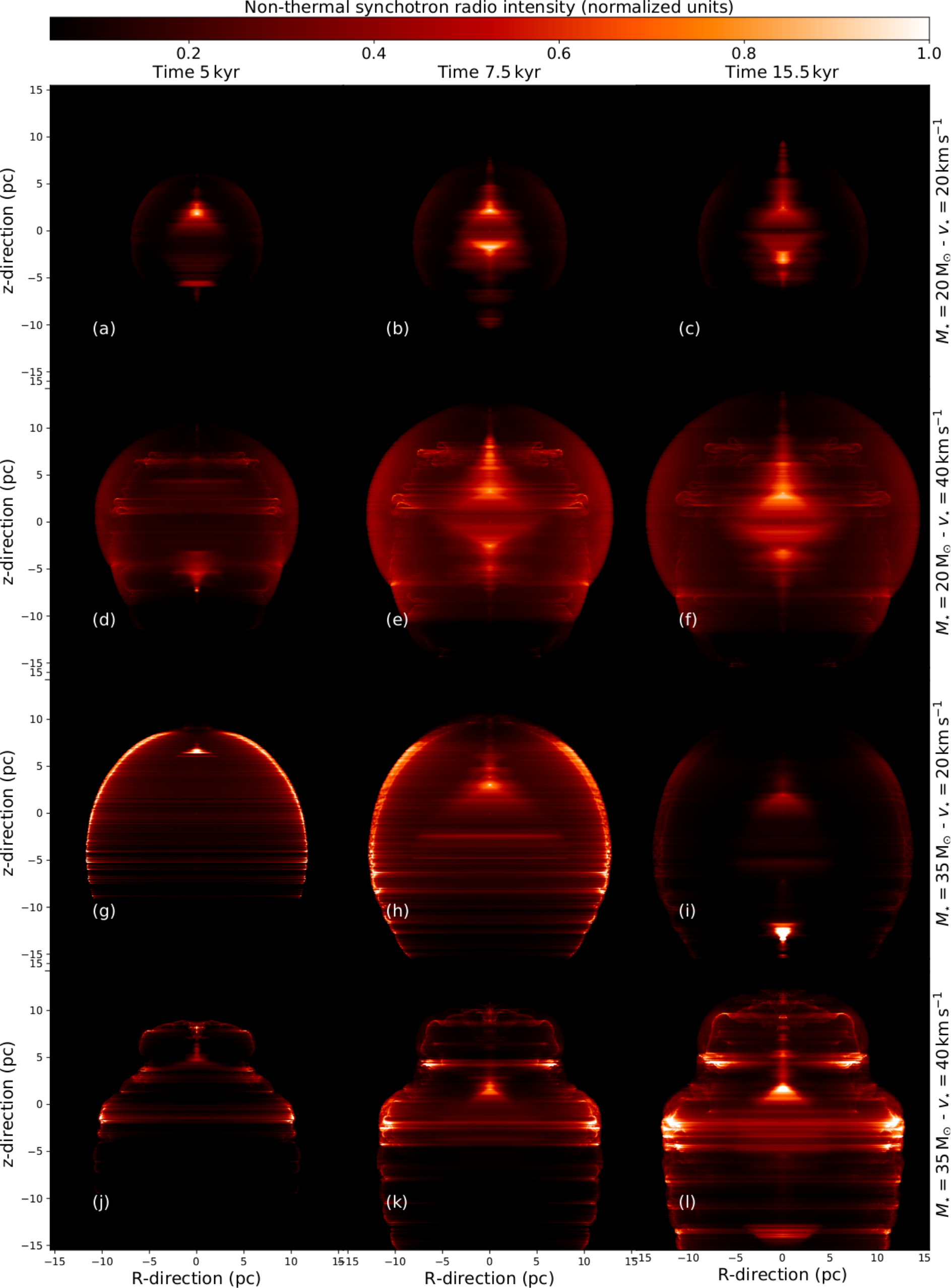}  \\        
        \caption{
        As Fig. \ref{fig:maps_sync_45deg} for $\theta_\mathrm{obs}=0^{\circ}$. 
        }
        \label{fig:maps_sync_0deg}  
\end{figure*}

Fig. \ref{fig:maps_sync_45deg} plots synchrotron emission maps of the plerionic 
supernova remnant models in this study, calculated assuming an aspect angle of 
$\theta_\mathrm{obs}=45^{\circ}$ between the observer's line-of-sight and the 
plane of the sky.
The figure displays the images as a function of time and of the initial conditions 
of the models, plotted in normalized units. 
Each column corresponds to a different 
time 
after the explosion, specifically $5\ \mathrm{kyrs}$ (left column), $7.5\ \mathrm{kyrs}$ (middle column), and $15.5\ \mathrm{kyrs}$ 
(right column), respectively. 
Each row corresponds to a different 
supernova progenitor, namely an initial mass $20\, \mathrm{M}_{\odot}$ moving with
velocity $v_{\star}=20\, \mathrm{km}\, \mathrm{s}^{-1}$ (top line), an 
initial mass $20\, \mathrm{M}_{\odot}$ moving with velocity 
$v_{\star}=40\, \mathrm{km}\, \mathrm{s}^{-1}$ 
(second row), an initial mass $35\, \mathrm{M}_{\odot}$ moving with 
velocity $v_{\star}=20\, \mathrm{km}\, \mathrm{s}^{-1}$ (third row), 
an initial mass $35\, \mathrm{M}_{\odot}$ moving with velocity 
$v_{\star}=40\, \mathrm{km}\, \mathrm{s}^{-1}$ (bottom row), respectively.
Fig. \ref{fig:maps_sync_0deg} is as Fig. \ref{fig:maps_sync_45deg} but assumes 
an aspect angle of $\theta_\mathrm{obs}=0^{\circ}$.

\begin{figure*}
        \centering
        \includegraphics[width=0.9\textwidth]{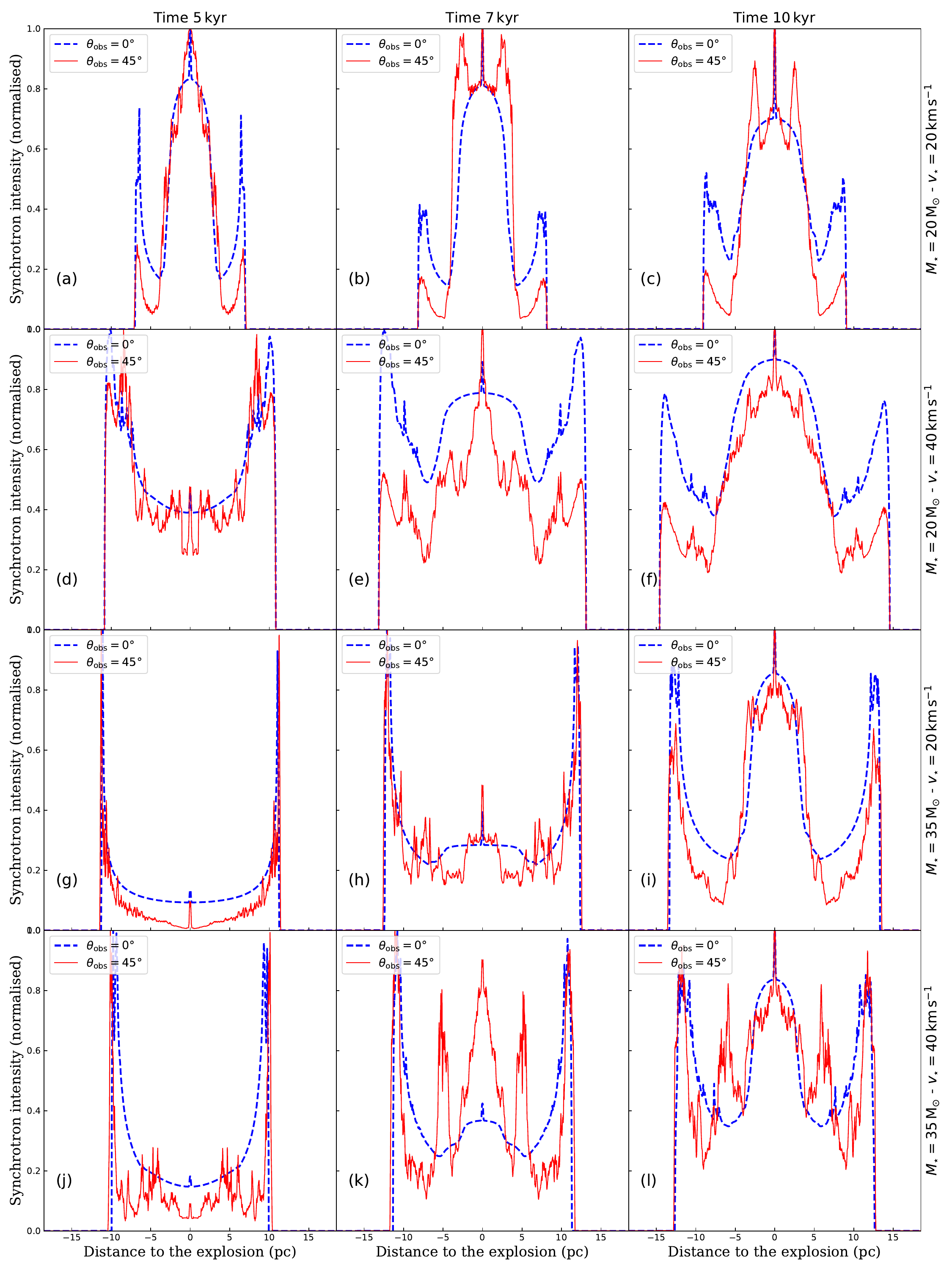}  \\        
        \caption{
        Cut through the synchrotron radio emission maps of the grid of models with 
        $\theta_\mathrm{obs}=45^{\circ}$, taken along the $Ox$ direction. 
        }
        \label{fig:cuts_SouthNorth_opt}  
\end{figure*}

\begin{figure*}
        \centering
        \includegraphics[width=0.9\textwidth]{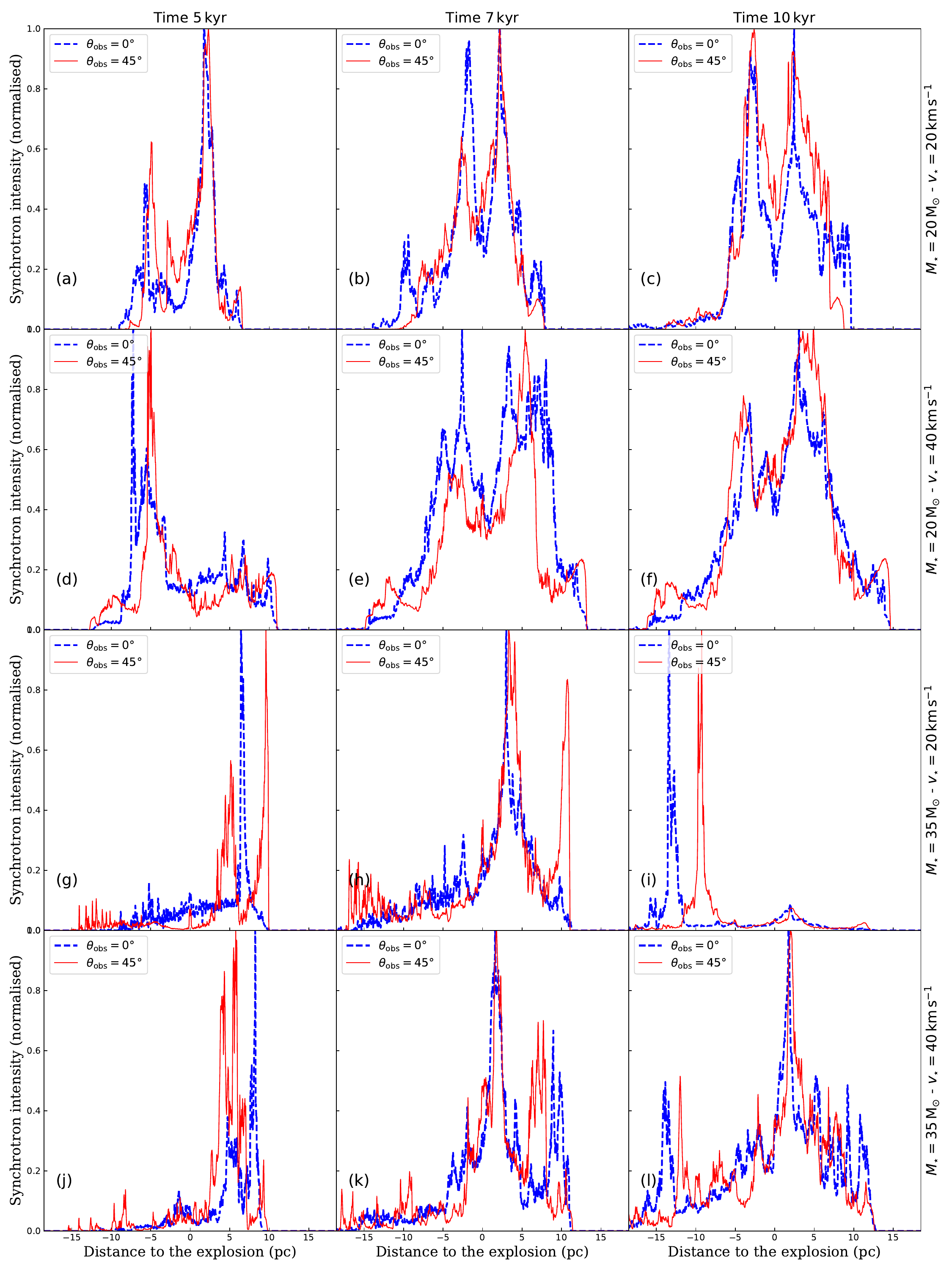}  \\        
        \caption{
        As Fig. \ref{fig:cuts_SouthNorth_opt} along the $Oy$ direction. 
        }
        \label{fig:cuts_SouthNorth_opt_2}  
\end{figure*}

In turn, Fig. \ref{fig:cuts_SouthNorth_opt} plots horizontal cuts taken through the synchrotron 
radio emission maps calculated with Fig. \ref{fig:maps_sync_0deg}, while 
Fig. \ref{fig:cuts_SouthNorth_opt_2} plots vertical cuts taken through the synchrotron 
radio emission maps calculated with $\theta_\mathrm{obs}=0^{\circ}$. 
The disposition 
of the different panels corresponds to that of Figs. \ref{fig:maps_sync_45deg}-\ref{fig:maps_sync_0deg}, 
comparing the emission for $\theta_\mathrm{obs}=0^{\circ}$ (dashed blue line) and 
$\theta_\mathrm{obs}=45^{\circ}$ (solid red line).
The second row of panels (Fig. \ref{fig:maps_sync_45deg} d,e,f) displays the maps 
already described in \citet{meyer_mnras_515_2022}. 
Initially, the pulsar wind is 
absent from the $z>0$ region that is along the direction of motion of its defunct 
progenitor star.
When the polar jet of the pulsar wind begins interacting with the supernova ejecta 
and the shocked stellar wind, it becomes brighter than the walls of the cavity 
shaped prior to the explosion (Fig. \ref{fig:maps_sync_45deg} e).

The north-south radio flux difference increases as a function of time: the northern 
jet collides into the shocked supernova material while the other one 
propagates in the cavity of low-density material, see Fig. \ref{fig:maps_sync_45deg}f.
The first row of panels corresponds to the model with a slower progenitor compared 
to that of the second line (Fig. \ref{fig:maps_sync_45deg} a,b,c). Therein, the
non-thermal flux from the circumstellar medium is much fainter than that of the 
pulsar wind, regardless of the time spent after the explosion and the onset of
the pulsar wind.
The north-south emission difference persists with a brighter northern part up to
$7.5\ \mathrm{kyrs}$ before being reversed at times $12.5\ \mathrm{kyr}$ when
the southern part becomes brighter (Fig. \ref{fig:maps_sync_45deg}c).

The model with velocity $35\, \mathrm{M}_{\odot}$ and moving with velocity 
$v_{\star}=20\, \mathrm{km}\, \mathrm{s}^{-1}$ induces a very bright circumstellar
medium that is much brighter than the pulsar wind nebula, appearing as a 
larger-scale synchrotron arc of radius $10\, \mathrm{pc}$ whose flux surpasses 
that of the jet.
At 
$7.5\ \mathrm{kyrs}$, the overall dense arc is even more luminous; however, 
the central pulsar wind nebula induces in its turn an increasing surface brightness
that competes with the region where the supernova blastwave interacts with the 
circumstellar medium (Fig. \ref{fig:maps_sync_45deg} g). 
The expanding collimated
wind of the spinning pulsar overwhelms the brightness of its surroundings made of
mixed ejecta, stellar wind, and ISM at time $12.5\ \mathrm{kyr}$ (Fig. \ref{fig:maps_sync_45deg} j),
as it is the case in the other models (Fig. \ref{fig:maps_sync_45deg} c,f).
The last model with a $35\, \mathrm{M}_{\odot}$ progenitor moving with velocity
$v_{\star}=40\, \mathrm{km}\, \mathrm{s}^{-1}$ has a circumstellar medium whose 
brightness is governed by a dense equatorial region where the ejecta meet the
unstable circumstellar medium, inducing bright synchrotron rings (Fig. \ref{fig:maps_sync_45deg} j). 
Throughout the entire modeled evolution of our plerion, the pulsar wind remains 
fainter than its surrounding supernova remnant, except in the very central region
where the polar originates, see Fig. \ref{fig:maps_sync_45deg}l.

\FloatBarrier

By 
comparing the emission maps at 
similar times, one notices that the region of the blastwave interacting with the circumstellar 
medium compresses the ISM magnetic field better in the case of a $35\, \mathrm{M}_{\odot}$ 
progenitor, screening the pulsar wind nebula; see images at time $5.0\ \mathrm{kyr}$ in
Figs. \ref{fig:maps_sync_45deg} a,d,g,j.
In the case of the $20\, \mathrm{M}_{\odot}$ progenitor, the interacting blastwave is brighter 
if the massive star moved faster ($v_{\star}=20\, \mathrm{km}\, \mathrm{s}^{-1}$, see 
Fig. \ref{fig:maps_sync_45deg} a,g), whereas in the case of the $35\, \mathrm{M}_{\odot}$ 
progenitor, the interacting region is brighter if the star moves slower 
($v_{\star}=40\, \mathrm{km}\, \mathrm{s}^{-1}$, Fig. \ref{fig:maps_sync_45deg} d,j).
At 
$7.5\ \mathrm{kyrs}$, the pulsar wind nebula shines brightly from its northern jet-like 
feature in all models, see Fig. \ref{fig:maps_sync_45deg} b,e,h,k. 
Major qualitative differences
between the simulation model happen at older times ($>10.0\ \mathrm{kyr}$), when the supernova
blastwave has interacted strongly with the circumstellar medium and is reverberating towards
the center of the explosion, interacting with the pulsar wind (Fig. \ref{fig:maps_sync_45deg} c,f,i,l).

\FloatBarrier

Fig. \ref{fig:maps_sync_0deg} displays the radio appearance of the plerion assuming an aspect 
angle of $\theta_\mathrm{obs}=0^{\circ}$. Obviously, the north-south surface brightness asymmetry 
does not exist when it is a direct result of the different angles between the observer's 
line-of-sight and the direction of the local magnetic field; see the model with 
$20\, \mathrm{M}_{\odot}$ moving with $v_{\star}=20\, \mathrm{km}\, \mathrm{s}^{-1}$ 
(Fig. \ref{fig:maps_sync_45deg}b).
This projection effect is also visible in the appearance of the polar cap in the model 
with $35\, \mathrm{M}_{\odot}$ when moving with velocity $v_{\star}=40\, \mathrm{km}\, \mathrm{s}^{-1}$, 
which is bright if $\theta_\mathrm{obs}=45^{\circ}$ but dark if $\theta_\mathrm{obs}=0^{\circ}$, 
respectively.
However, several time instances still display such features when it comes from the changes
in gas number density when the jet interacts with the surrounding medium. This particularly 
applies to the images in Fig. \ref{fig:maps_sync_45deg}e,h,k, at times $7.5\ \mathrm{kyr}$.

The rings in the emission maps are a consequence of the 3D reconstruction of the density 
and magnetic structures on the basis of 2.5D simulations, when bright clumps
(Fig. \ref{fig:maps_sync_0deg} h) and Rayleigh-Taylor instabilities (Fig. \ref{fig:maps_sync_45deg} e) 
of the ejecta/wind discontinuity are rotated around the symmetry axis of the models.
The cross-sections taken through the emission maps reveal that the remnants with 
$20\, \mathrm{M}_{\odot}$ are brighter when observed with an aspect angle of 
$\theta_\mathrm{obs}=45^{\circ}$ compared to that of $\theta_\mathrm{obs}=0^{\circ}$, 
see Fig. \ref{fig:cuts_SouthNorth_opt}a,d and see Fig. \ref{fig:cuts_SouthNorth_opt}b,e.
This is 
not true in the case of the $35\, \mathrm{M}_{\odot}$ progenitor, 
whose models have roughly 
the same surface brightness regardless of the considered 
aspect angle, except at some specific time instances, see Fig. \ref{fig:cuts_SouthNorth_opt}g.
The horizontal cross-sections display a more complex pattern if $\theta_\mathrm{obs}=45^{\circ}$
compared to $\theta_\mathrm{obs}=0^{\circ}$, see, e.g., Fig. \ref{fig:cuts_SouthNorth_opt} k,l.

The vertical surface brightness slices further illustrate the respective positions of the 
emission peaks, corresponding to the jets of the pulsar wind in the model with
$20\, \mathrm{M}_{\odot}$ see Fig. \ref{fig:cuts_SouthNorth_opt_2}a-c and 
Fig. \ref{fig:cuts_SouthNorth_opt_2}d-f. 

The respective intensity of the radio 
emission changes as a function of the inclination of the plane of the object with
respect to the plane of the sky (Fig. \ref{fig:cuts_SouthNorth_opt_2}b,c).
In the simulations with $35\, \mathrm{M}_{\odot}$ the relative intensity of the polar 
jet and the interacting blastwave in the supernova remnant is more pronounced if 
$\theta_\mathrm{obs}=0^{\circ}$ compared to when $\theta_\mathrm{obs}=45^{\circ}$,
see Fig. \ref{fig:cuts_SouthNorth_opt_2}h.
Again here, the surface brightnesses in the models with a $35\, \mathrm{M}_{\odot}$ progenitor 
moving with velocity $v_{\star}=40\, \mathrm{km}\, \mathrm{s}^{-1}$ are roughly 
the same, regardless of the aspect angle under which they are considered.

\section{Discussion}
\label{discussion}

This section presents the limitation of the method and 
compare the results with observations.

\subsection{Limitations of the models}
\label{limitations}

As already reported in the preceding studies using this numerical approach, 
see \citet{meyer_mnras_515_2022,meyer_527_mnras_2024}, there are two key 
caveats inherent to this type of simulations. 
First, it is imperative to 
understand that the simulations conducted within this study are confined to a 
two-dimensional framework, assuming axisymmetry, thereby disregarding any 
potential variations in the supernova progenitor's axis of rotation and/or 
the pulsar's spin relative to the direction of motion of the massive star and/or 
the direction of the ISM magnetic field. This is then an exploratory work. 
While our current approach enhances 
computational efficiency and provides 
insights into the supernova 
remnant problem, only a full three-dimensional treatment would  
be realistic, permitting, for example, the inclusion 
of factors such as magnetic turbulence and the multi-phased structure of the 
ISM, which could profoundly influence the development of the stellar wind 
bubble and the expansion of the supernova blast wave.
Consequently, it becomes evident that future investigations employing a three-dimensional 
approach hold the promise of offering a more comprehensive understanding of these 
complex phenomena. However, the computation cost of such research needs to be downsized before they are deemed possible.

Furthermore, an additional aspect that warrants meticulous 
consideration is the absence of pulsar motion in the simulations. Integrating 
pulsar motion into our models would introduce a layer of complexity that is 
currently absent, thereby affording a more realistic representation of the 
intricate interplay between the pulsar wind and its surrounding medium. 
Moreover, 
accounting for the oblique rotation of the pulsar's magnetic axis would 
facilitate a more accurate reproduction of the observed characteristics of 
pulsar wind nebulae, e.g., by both non-thermal synchrotron and inverse 
Compton radiation mechanisms. 
These aforementioned aspects underscore 
promising avenues for future research.

It is also to note that the pulsar itself does not exhaust the possible features
of these systems. We have assumed a fixed initial spin-down power, a fixed initial period and period derivative, a fixed wind 
magnetization. There is no radiative model incoroporated for the pulsar electron population itself.

\subsection{Concluding remarks on emission maps}
\label{resultats_emission_maps}

Taking into account the limitations, the series of emission maps presented in this study cover the parameter space of the most 
common core-collapse supernova progenitors and explore the most probable bulk velocities 
reached when animated by a supernova moving through the ISM.  
Our models therefore measure the deviations from the barrel-to-Cygnus loop evolutionary 
sequence, as identified in \citet{2024arXiv240407873M}, that are induced by the presence 
of a pulsar wind blown after the death of the progenitor. 
A few global remarks can be made. The synchrotron emission maps of the initial 
$20\, \mathrm{M}_{\odot}$ (red supergiant) progenitor form a bright jet-like feature, 
which is surrounded, in the case of a high space velocity of the star 
($v_{\star}=40\, \mathrm{km}\, \mathrm{s}^{-1}$), by a fainter arc embedding the central 
pulsar wind nebula. In other words, the pulsar wind nebula is the brightest component of 
the supernova remnant. In contrast, for the $35\, \mathrm{M}_{\odot}$ (Wolf-Rayet) 
progenitor, the surrounding arc resulting from the interaction of the blast wave with the 
circumstellar medium emits more than the central pulsar wind nebula, regardless of the 
inclination angle of the axis of the remnant with respect to the plane of the sky 
(see Figs. \ref{fig:maps_sync_45deg}, \ref{fig:maps_sync_0deg}).  
This implies that an overhanging surrounding shell-like feature around a synchrotron 
pulsar wind nebula is a necessary ingredient to identify before associating its 
progenitor to an initial mass higher than $\sim 20\, \mathrm{M}_{\odot}$. 

The cuts taken laterally through the pulsar wind nebulae show that the models with a 
progenitor mass of $20\, \mathrm{M}_{\odot}$ (red supergiant) are centrally peaked, 
especially at later times ($>7\, \rm kyr$), regardless of the inclination of the remnant 
with respect to the plane of the sky, see Fig \ref{fig:cuts_SouthNorth_opt}b, c, e, f. 
When the cross-sections are considered along the direction of motion of the progenitor 
star, two peaks are clearly noticeable if the progenitor mass is $20\, \mathrm{M}_{\odot}$, 
except in the early pulsar wind nebula formation phase (Fig \ref{fig:cuts_SouthNorth_opt_2}d). 
The situation is more complex and irregular in the case of the supernova remnants 
generated by a $35\, \mathrm{M}_{\odot}$ (Wolf-Rayet) star, in the sense that a 
single intensity peak forms at the forward shock at the beginning of the pulsar 
wind expansion, and a second peak forms later. 
Finally, atypical patterns develop in the context of a fast-moving progenitor 
($v_{\star}=40\, \mathrm{km}\, \mathrm{s}^{-1}$), see Fig \ref{fig:cuts_SouthNorth_opt_2}k, l, 
making it difficult to identify  
characteristic synchrotron maps. This is 
due to the larger cavity shaped by the stellar wind prior to the explosion, inducing 
a complex reverberation of the shock wave and of the pulsar wind termination shock. 

\subsection{Comparison with particular pulsar wind nebulae}
\label{potential_progenitor}


The initial mass function (IMF) of \citet{kroupa_mnras_322_2001} indicates that the 
distribution of zero-age main-sequence star masses in clusters follows a 
power-law decrease. This means that most core-collapse supernova progenitors are in 
the red supergiant phase at the moment of their explosion, while others are in 
the Wolf-Rayet evolutionary phase \citep{katsuda_apj_863_2018}. 
%
%
Despite the fact that a significant proportion of Wolf-Rayet stars directly 
collapse into black holes and generate gravitational waves rather than producing 
neutron stars \citep{2017A&A...603A.120V,2018NewA...58...33B}, the details of 
the mechanisms distinguishing these two scenarios are not fully 
understood \citep{2011MNRAS.414.2985D}. The constraint of identifying a supernova 
progenitor based on its supernova remnant and/or pulsar wind nebula is 
still not possible on a systematic basis \citep{2022Natur.601..201G}. 
However, a few individual pulsar wind nebulae can be directly compared to our simulations. 

%
The Jellyfish nebula around the pulsar B1509–58 in the supernova remnant 
G320.4–1.2/MSH 15–52 forms a complex structure that has been studied 
at various wavelengths. It reveals outflows expanding asymmetrically into 
a local medium that is, in some regions, of low density. This suggests a 
governing role of a circumstellar cavity in shaping the pulsar wind 
nebula \citep{1991AJ....101.2160A,2002ASPC..271..175G,2002AJ....123..337D,
2010ApJ...714..927A,2022ApJ...927...87H}. 
The discovery of a series of open rings surrounding the pulsar, interpreted as a 
(potentially reverberating) termination shock \citep{2009PASJ...61..129Y}, 
is consistent with our simulations. In the early formation 
phase of the pulsar wind nebula, its morphology along the direction of motion 
of the progenitor is round 
instead of adopting the 
typical X-shape described by \citet{komissarov_mnras_349_2004}, see 
Figs. \ref{fig:maps_sync_45deg}a, \ref{fig:maps_sync_0deg}a. 
This leads us to think that the pre-supernova circumstellar medium of 
G320.4–1.2/MSH 15–52 was asymmetric, perhaps induced by the bulk motion 
of the supernova progenitor through its ambient medium, and that the peculiar 
morphology of G320.4–1.2/MSH 15–52 is partly due to this same mechanism that 
we explore in the present study. 

%
The plerionic supernova remnant N158A and its pulsar B0540–69 
\citep{2007ApJ...662..988P,2011MNRAS.413..611L}. N158A is an 
extragalactic remnant in the Large Magellanic Cloud with a progenitor mass estimated 
to be about $20-25\, \rm M_\odot$, although the O and S lines in its enriched environment 
revealed by X-ray spectrometry suggest a heavier, Wolf-Rayet progenitor 
\citep{2008ApJ...687.1054W}. This may not necessarily be inconsistent, as the 
zero-age main-sequence mass of Wolf-Rayet stars can be smaller at low metallicity 
than in the Milky Way \citep{2014A&A...565A..27H}. 
It displays a structure made of a jet-like feature and an equatorial disc-like 
with little evidence of interaction 
with the surrounding circumstellar medium. This is consistent with the Wolf-Rayet 
progenitor evolutionary scenario, with a large cavity carved by a powerful 
wind, see Fig. \ref{fig:snr_pwn_3520}a,b and Fig. \ref{fig:snr_pwn_3540}a,b. 
\section{Conclusion}
\label{conclusion}

This study explores the effects of a pulsar wind nebula onto the 
shaping of plerionic supernova remnants generated by runaway massive stars, 
within an explored parameter space corresponding to the most common 
supernova remnants, both in terms of progenitor mass and bulk motion through the ISM. 
By means of 2.5 magneto-hydrodynamical simulations 
(2 dimensions for the scalar quantities plus a toroidal component for the vectors), we first model the circumstellar 
medium of $20$ and $35\, \rm M_{\odot}$ stars moving with velocities $20$ and 
$40\, \rm km\, \rm s^{-1}$, in which we deposit the blastwave of a core-collapse 
supernova remnant \citep{truelove_apjs_120_1999}, followed by the wind of a 
rotating neutron star \citep{vanderswaluw_aa_404_2003,komissarov_mnras_349_2004}, 
that is advanced from its young to middle-age ($12.5\, \rm kyr$) evolution 
times \citep{meyer_mnras_515_2022}. 
\textcolor{black}{
Within this approach, the direction of motion of the runaway massive star, 
the direction of the local ambient medium magnetic field assumed to be organised, 
the axis of rotation of the progenitor as well as that of the pulsar are all aligned. 
This imposes our models to apply to a peculiar class of pulsar wind nebulae, 
driven by high equatorial energy flux.
}  
The progenitors are considered to live and move into the warm phase of the galactic 
plane where the background number density amounts $n_{\rm ISM}=0.79\, \rm cm^{-3}$ 
for an equilibrium temperature of $T_{\rm ISM}=8000\, \rm K$, where they die and 
form plerionic remnants. 
The simulations are performed with the well-tested astrophysical code \textsc{pluto} \citep{mignone_apj_170_2007,migmone_apjs_198_2012} using the methodology 
developed for core-collapse supernova remnants \citep{meyer_mnras_493_2020,
meyer_mnras_502_2021,meyer_mnras_521_2023} and for pulsar wind nebulae \citep{meyer_mnras_515_2022,meyer_527_mnras_2024}. 
The outputs are post-processed for radio synchrotron emission maps to be discussed 
in the context of real observations, see the studies of \citet{velazquez_mnras_519_2023,
villagran_mnras_527_2024}.

Our models show that the distribution of circumstellar material prior 
to the supernova explosion is relevant to 
the reverberation of the termination 
shock of the pulsar wind nebula \citep{Bandiera_mnras_165_2023,2023_mnras_2839_2023}, 
inducing morphologies which deviate from the classical diamond-like 
solution of \citet{komissarov_mnras_349_2004}. 
This is particularly pronounced when the supernova explosion takes place 
deeply embedded into the wind bubble of the progenitor (slowly-moving 
massive star) or when the distribution of pre-supernova circumstellar 
medium is particularly dense (high-mass progenitor such as Wolf-Rayet 
star). Conversely, it is less pronounced when the explosion happens 
out of the wind bubble, i.e. in the context of a Cygnus Loop remnant 
\citep{aschenbach_aa_341_1999} produced by a fast-moving red supergiant 
star \citep{brighenti_mnras_270_1994}.  
Those asymmetric pulsar wind nebulae display a large variety of non-thermal 
radio projected emission, either screened by the overhanging interacting 
supernova blastwave or via projected asymmetric up–down synchrotron emission  
\citep{meyer_mnras_515_2022}. 
The reverberation of the pulsar reverse shock plays a role in a wider 
mechanism that is the reverberation of the supernova shock wave towards 
the center of the explosion. The pulsar wind prevents it from happening 
entirely and limits the back-and-forth reflection of core-collapse 
blastwaves inside of their progenitor's circumstellar medium 
\citep{dwarkadas_HEDP_9_2013,fang_mnras_464_2017}. 
This process must have important consequences onto the mixing of materials 
inside of plerionic supernova remnants, which we hope to investigate in the future. 
\textcolor{black}{
Another aspect that we are currently addressing is the development of full three-dimensional 
simulations. These models will better account for magneto-hydrodynamic processes, such as the 
instabilities that significantly influence the dynamics and morphology of the flow.  
Particularly, the question of the jet originating from a moving pulsar and interacting 
with the circumstellar medium will then be tackled. 
} 
\textcolor{black}{
This will also remove the constraint imposing the alignment of the different rotation axises and 
direction of motion involved in this problem, greatly enhancing the degree of realism of the 
models. 
}


\section*{Acknowledgements}

\textcolor{black}{
The author thankfully acknowledges RES resources provided by 
BSC in MareNostrum to AECT-2024-2-0002. 
}
The authors gratefully acknowledge the computing time made available to them 
on the high-performance computer "Lise" at the NHR Center NHR@ZIB. This 
center is jointly supported by the Federal Ministry of Education and 
Research and the state governments participating in the NHR 
(www.nhr-verein.de/unsere-partner). 
This work has been supported by the grant PID2021-124581OB-I00 funded by 
MCIN/AEI/10.13039/501100011033 and 2021SGR00426 of the Generalitat de Catalunya. 
This work was also supported by the Spanish program Unidad de Excelencia Mar\'ia 
de Maeztu CEX2020-001058-M. 
This work also supported by MCIN with funding from European Union 
NextGeneration EU (PRTR-C17.I1).

\section*{Data availability}

This research made use of the \textsc{pluto} code developed at the University of Torino 
by A.~Mignone (\url{http://plutocode.ph.unito.it/}).
and of the \textsc{radmc-3d} code developed at the University of Heidelberg by C.~Dullemond 
(\url{https://www.ita.uni-heidelberg.de/~dullemond/software/radmc-3d/}).
The figures have been produced using the Matplotlib plotting library for the 
Python programming language (\url{https://matplotlib.org/}).
The data underlying this article will be shared on reasonable request to the 
corresponding author.


\bibliographystyle{aa} 
\bibliography{grid} 


\end{document}